\documentclass{IEEEtran}
\usepackage{cite}
\usepackage{amsmath,amssymb,amsfonts}
\usepackage{algorithmic}
\usepackage{graphicx}
\usepackage{textcomp}
\def\BibTeX{{\rm B\kern-.05em{\sc i\kern-.025em b}\kern-.08em
    T\kern-.1667em\lower.7ex\hbox{E}\kern-.125emX}}

\usepackage{pdfpages}
\usepackage{threeparttable} 
\usepackage{multirow} 
\usepackage{booktabs}
\usepackage{enumerate}
\usepackage{algorithm}
\usepackage{algorithmic}
\usepackage{epsfig}
\usepackage{multicol} 
\usepackage{stfloats}

\newtheorem{theorem}{Theorem}
\newtheorem{proposition}{Proposition}
\newtheorem{lemma}{Lemma}

\newtheorem{corollary}{Corollary}
\newtheorem{example}{Example}
\newtheorem{remark}{Remark}

\newtheorem{assumption}{Assumption}
\newenvironment{proof}{\hspace{0ex}\textsc{Proof}.\hspace{1ex}}{\hfill$\Box$\newline}
\makeatletter

\newcommand{\Rmnum}[1]{\expandafter\@slowromancap\romannumeral #1@}
\usepackage{setspace}                                    

\begin{document}
\title{Performance Analysis of Distributed  Filtering under  Mismatched  Noise Covariances}
\author{Xiaoxu Lyu, Guanghui Wen, Ling Shi, Peihu Duan, and Zhisheng Duan

\thanks{
Xiaoxu Lyu and Ling Shi are with the Department of Electronic and Computer Engineering, Hong Kong University of Science and Technology, Hong Kong, China (e-mail: eelyuxiaoxu@ust.hk; eesling@ust.hk).}

\thanks{Guanghui Wen is with the Laboratory of Security Operation and Control for Intelligent Autonomous Systems, Department of Systems Science, School of Mathematics, Southeast University, Nanjing 211189, China (e-mail: wenguanghui@gmail.com).}

\thanks{Peihu Duan is with School of Electrical Engineering and Computer Science, KTH Royal Institute of Technology, Stockholm, Sweden (e-mail: peihu@kth.se).}

\thanks{
Zhisheng Duan is with State Key Laboratory for Turbulence and Complex Systems, Department of Mechanics and Engineering Science, College of Engineering, Peking University, Beijing 100871, China (e-mail: duanzs@pku.edu.cn).}

}

\maketitle

\begin{abstract}
This paper systematically  investigates the performance of consensus-based distributed  filtering under mismatched noise covariances.
First,   we introduce   three  performance evaluation indices for such filtering problems, namely the standard performance evaluation index, the nominal performance evaluation index, and the estimation error covariance.
We  derive  difference expressions among these indices and establish  one-step relations among them  under various mismatched noise covariance  scenarios.  We particularly reveal the effect of the consensus fusion on these relations. Furthermore, the recursive relations   are introduced  by extending the results of the one-step relations.
 Subsequently, we demonstrate the convergence  of these indices  under the collective observability condition, and show this convergence condition of the nominal performance evaluation index can guarantee the convergence of the estimation error covariance. Additionally, we prove that the estimation error covariance of the consensus-based distributed filter under mismatched noise covariances can be bounded by the Frobenius norms of the noise covariance deviations and the trace of the nominal performance evaluation index. Finally, the effectiveness of the theoretical results is verified by numerical simulations.
 %
\end{abstract}


\section{Introduction}

%
%
%
%
%
%
%

Over the past  two decades,  sensor networks
 have emerged as a research focus within the systems and control community, partly  due to its  wide  applications  in the fields of wide-area monitoring  \cite{xie2012fully}, target tracking \cite{meyer2017scalable},  cooperative mapping \cite{gao2020random}, etc.
%
According to whether there is a central node managing the whole network,   state estimation
 methodologies of sensor networks can be
  generally  classified into two categories:
 centralized state  estimation  and distributed state  estimation.
In  comparison  to  centralized state estimation, which is characterized by  high communication overheads, substantial computational burdens,  and vulnerability to  failures of  the center node,
 distributed state estimation presents considerable advantages, including  low computation costs, light computation burdens, and enhanced reliability.
   Specifically,  distributed state estimation is aimed to estimate the state of the target system for each sensor in  sensor networks by using only its local information.


The advancement of consensus theory for multi-agent systems \cite{olfati2004consensus1,olfati2007consensus,li2009consensus,qin2016recent}
has  provided  insights into the distributed state estimation problem. Consequently, quite  a few
 consensus-based  distributed state estimation approaches have been proposed \cite{olfati2007distributed,cattivelli2010diffusion,battistelli2014kullback,
battistelli2014consensus,he2018consistent,duan2020distributed,qian2022consensus,duan2022distributed}. By utilizing the consensus-based techniques,  the local information, including local measurements\cite{olfati2007distributed,qian2022consensus}, local information matrices \cite{battistelli2014kullback,he2018consistent}, or local state estimates\cite{duan2022distributed}, can be collected in a distributed manner  and fused by each sensor.
 As a  pioneering work,
Olfati-Saber \cite{olfati2007distributed} proposed a distributed  Kalman filter   by utilizing two consensus filters to fuse  measurements and covariance information.
Cattivelli and Sayed \cite{cattivelli2010diffusion} proposed
a diffusion Kalman filtering strategy for distributed state estimation  to handle  the filtering and smoothing problems.
Then, by exploiting the Kullback-Leibler average consensus of the local probability density functions,  Battistelli and Chisci \cite{battistelli2014kullback} derived
a distributed state estimation algorithm.  Later, by combining the consensus on measurements and   the consensus on information matrices, Battistelli et~al. \cite{battistelli2014consensus}  presented
a hybrid consensus-based filter. Furthermore,  Duan et~al. \cite{duan2022distributed}  proposed  a  distributed estimation framework  for a continuous-time  system by using a leader-following consensus   information fusion  strategy.

The aforementioned results  provide  numerous  options  for  distributed state estimation  in engineering applications.
However, it is  important to note that these results are derived based on the accurate models.
In practical scenarios, it is common that  noise covariances are mismatched   due to modeling errors  and  environmental disturbances. In other words,   there usually exist deviations between the nominal noise covariances and the actual noise covariances.
The presence of  mismatched noise covariances may lead to performance degradation and potentially destroy the stability of the filter.
 For this tricky issue, some relevant researches have been made.
  A  viable  solution to address this issue involves utilizing
    adaptive strategies  to estimate the unknown covariances \cite{mehra1970identification,mehra1972approaches,karasalo2011optimization,yu2012ins}.
Specifically,  Mehra \cite{mehra1970identification} proposed a
  scheme to identify unknown noise covariances. Later,  Mehra \cite{mehra1972approaches} introduced
  four  methods of handling the effect of unknown noise covariances, namely Bayesian, maximum likelihood, correlation, and covariance matching approaches.  Karasalo and Hu  \cite{karasalo2011optimization}  were  concerned with the optimization-based adaptive Kalman filtering method.
In more recent years,  the variational  Bayesian  method   has been  used to solve this problem by  simultaneously  estimating  the states and the  unknown noise covariances, which are usually designed as the specific  conjugate prior distributions
\cite{sarkka2009recursive,huang2017novel,lv2022stochastic}.

  Although some  methods to handle unknown noise covariances   have been proposed,  the  performance  of the distributed filters  in the presence of   mismatched noise covariances has not  been thoroughly analyzed.  To be specific,
the convergence  and  the performance degradation due to the presence of  mismatched noise covariances have not been  quantitatively investigated.
In the context of state estimation for  single-sensor systems, a body of relevant literature exists \cite{fitzgerald1971divergence,toda1980bounds,sangsuk1990analysis,ge2016performance,shi2007kalman}.
Specifically,  Fitzgerald \cite{fitzgerald1971divergence} showed that the mean-square estimation errors of the filter  may diverge due to the increasing intensity of the process noise. Toda and Patel \cite{toda1980bounds} derived
the bounds on  the estimation errors in the presence of noise covariance errors. Sangsuk-Iam and Bullock \cite{sangsuk1990analysis} analyzed the convergence and divergence properties  of the discrete-time Kalman filtering. Recently,
Ge et al. \cite{ge2016performance}  utilized  three mean squared errors, including the ideal, the calculated,  and the true mean squared errors,
to  evaluate  the  performance  of the  filter under   mismatched noise covariances. For sensor networks, Shi et~al.  \cite{shi2007kalman} investigated the filtering  algorithm using centralized sensor fusion under uncertain noise covariances.
However, there  still lacks  systematic  analysis for  distributed filtering under mismatched noise covariances.

Regardless of whether there are   mismatched covariances or not,
the convergence    is an important  metric  for evaluating the performance of  distributed filters. There exists some literature focusing on the convergence analysis   under accurate noise covariances by utilizing the Riccati equations.
Callier and Willems  \cite{callier1981criterion} provided
 a criterion for the convergence of the solution of the Riccati differential equation.  Chan et al.~\cite{chan1984convergence} investigated
 the convergence properties  of the Riccati difference equation of    nonstabilizable systems.  Then, Nicolao~\cite{de1992time} analyzed
the time-varying Riccati difference equation, which played a  fundamental role in understanding  convergence and  stability. Li et al. \cite{li2015distributed} proposed
  a modified gossip interactive Kalman filter, and  analyzed a tradeoff between the communication rate and the  estimation performance with the random Riccati equation.
 Qian el al. \cite{qian2022consensus} established the performance gap  between  the consensus-based  distributed filter and  the centralized filter by introducing a modified algebraic Riccati equation.    However, these techniques may be ineffective for analyzing the performance of distributed  filters under mismatched noise covariances, since the actual estimation error covariances can not be  evaluated by  the nominal  algebraic  Riccati equations.

Motivated by the above observations, this paper aims to
provide a systematic  investigation on the performance of the distributed filtering under mismatched noise covariances.
The main contributions of this paper are summarized below:
\begin{enumerate}
\item
To  systematically   investigate the performance  of the distributed filtering under mismatched noise covariances,
we  introduce   three  performance evaluation indices, namely  the standard performance evaluation index, the nominal performance evaluation index, and the estimation error covariance. Accordingly, the difference expressions between them are formulated (\textbf{Propositions \ref{lemma sigmat-sigma}-\ref{lemma sigmaf-sigma}}).  These difference expressions play an essential role in analyzing the relations among different performance evaluation indices and  shedding light on the effect of the fusion step $L$  on these relations.

\item  The one-step relations among these performance evaluation indices, defined as  the $k$-step relations  provided that they are equal at the $(k-1)$-step,  are discussed (\textbf{Theorems~\ref{theorem phi q=0},~\ref{theorem q r}}). Then, the recursive relations are given by extending the results  of the one-step relations (\textbf{Theorem~\ref{theorem recurse}}). In addition,  we  reveal  the effect of  the fusion step $L$ on  the relations among these performance evaluation indices (\textbf{Theorem~\ref{lemma katamata inequality}}).

\item  Under the collective  observability  condition,   the nominal performance evaluation index and the estimation error covariance converge to the unique solutions of the algebraic Riccati equation  and the Lyapunov equation, respectively (\textbf{Theorem~\ref{eq therotem convergence}}).
    The convergence of the estimation error covariance is guaranteed by the  convergence  condition of the nominal system. Then,  based on two  steady-state solutions,  the trace of  the estimation error covariance  is bounded by using
    the Frobenius norms of the mismatched noise covariance deviations and the trace of the nominal performance evaluation index
     (\textbf{Theorem~\ref{theorem sigmaf sigmat bound}}).
\end{enumerate}

The remainder of this paper is organized as follows. Preliminaries, including graph concepts and some useful lemmas, and the problem formulation  are introduced in Section~\ref{sec pre and problem fomulation}. Three  performance evaluation indices under mismatched noise covariances  are presented in Section~\ref{section different estimation error covariance}.
The difference expressions  among different performance evaluation indices are derived, and relations among these indices are analyzed in Section~\ref{sec analyze difference }. The convergence of two performance evaluation indices is proven, and
 the bounds of the performance of the asymptotic distributed filter are given by  utilizing two solutions of the corresponding equations in Section~\ref{sec convergence analysis}. The obtained results are validated by numerical simulations  in Section~\ref{sec simulation}. Finally, conclusions are drawn in Section~\ref{sec conclusion}.

\textit{Notations}:  Throughout this paper, let $\mathcal{R}^n$ and $\mathcal{R}^{n\times n}$ be the sets of $n$-dimensional real vectors and $n\times n$-dimensional real matrices, respectively.
For a matrix $A\in \mathcal{R}^{n\times n}$, $A^{-1}$ and $A^{T}$ represent its inverse and transpose, respectively,
$||A||_F$ is the Frobenius  norm,  $\rho(A)$ is the spectral radius, and $\sigma(A)$ is the maximum singular value. Notation $\circ$ denotes the Hadamard product, and $\otimes$  represents the Kronecker product.
Symbol $\text{vec}(A)$ represents  the column vector composed   by concatenating the columns of the matrix $A$, and $[A]_{ij}$ denotes the $(i,j)$-th element of the matrix $A$.  The matrix inequalities $A > B$  and  $A\geq B$ indicate that $A-B$ is positive definite and positive semi-definite, respectively.  $E\{x\}$ refers to the expectation of the random variable $x$,  and $\text{Tr}(A)$  represents the trace of the matrix $A$.

\section{Preliminaries and Problem Statement}\label{sec pre and problem fomulation}

\subsection{Graph Theory}

 A sensor network is denoted by a communication topology ${\mathcal{G}}(\mathcal{V},\mathcal{E},\mathcal{L})$, which is composed of a communication node set $\mathcal{V}=\{1,2,\ldots,N\}$, an edge set $\mathcal{E} \subseteq \mathcal{V} \times \mathcal{V}$, and an adjacency matrix $\mathcal{L}=[l_{ij}]$.  The symbol $l_{ij}>0$  represents that sensor $j$ can  transmit information to sensor $i$. Thus, sensor $j$ is called the in-neighbor of sensor $i$. The  in-neighbor set of sensor $i$ is denoted as $\mathcal{N}_i$. A doubly stochastic matrix $\mathcal{L}$ means $\sum^N_{j=1}l_{ij}=\sum^N_{i=1}l_{ij}=1$. In addition,
 an undirected  graph  has the property that $(j, i)\in \mathcal{E}$ implies  $(i, j)\in \mathcal{E}$. A graph is called connected, if there exists a path for any pair of distinct nodes.

\subsection{Some Useful Lemmas}

\begin{lemma}(Matrix Inversion Lemma)\label{lemma matrix inverse lemma}
For matrices  $A$, $B$, $C$, and $D$ with proper dimensions, if $A^{-1}$ and $C^{-1}$ exist, then
\begin{equation*}
\begin{aligned}
(A + BCD)^{-1} = A^{-1}- A^{-1}B(C^{-1}+DA^{-1}B)^{-1}DA^{-1}.
\end{aligned}
\end{equation*}
\end{lemma}

\vspace{6pt}
\begin{lemma}\label{lemma a+baa+b}
For matrices $A$ and $B$ with proper dimensions and a constant number $a$, if $A^{-1}$ and $B^{-1}$ exist, then
\begin{equation*}
\begin{aligned}
&~~~~a(A+B)^{-1}A(A+B)^{-1}-(A+B)^{-1}~~~\\
& =(A+B)^{-1}((a-1)I-aB(A+B)^{-1}).
\end{aligned}
\end{equation*}
\end{lemma}

The proof of Lemma \ref{lemma a+baa+b} is given in  Appendix \ref{appdex lemma 2}.

\vspace{6pt}
\begin{lemma}\label{lemma a+b-a}
For matrices $A$ and $B$ with proper dimensions, if $A^{-1}$ and $B^{-1}$ exist, then
\begin{equation*}
\begin{aligned}
(A+B)^{-1}-A^{-1} =  -A^{-1}B(A+B)^{-1}.
\end{aligned}
\end{equation*}
\end{lemma}

The proof of Lemma \ref{lemma a+b-a} is given in  Appendix \ref{appdex lemma a+b-a}.

%
%
%
%
%
%

\vspace{6pt}
\begin{lemma}\cite{qian2022consensus}\label{lemma lapalacian metrix}
If the communication topology is strongly connected and  the corresponding matrix  $\mathcal{L}$ is  doubly stochastic with positive diagonal elements, then the matrix $\mathcal{L}^m$ converges to $\frac{1}{N}11^T$ exponentially as $m \to \infty$.

\end{lemma}

\vspace{6pt}
\begin{lemma}\cite{sinopoli2004kalman}\label{lemma k min}
For matrices $K$, $H$, $\Sigma$, and $R$ with proper dimensions, suppose that $\Sigma>0$, $R>0$, and $K$  is a variable. Let $\psi(K)= (I-KH)\Sigma(I-KH)^{T}+KRK^T$ and $K_{\text{min}}=\Sigma H^T(H\Sigma H^T+R)^{-1}$. Then,   $\psi(K)$ is quadratic and convex in the variable $K$. Moreover,  $\psi(K_{\text{min}})\leq \psi(K)$ for all $K$.
\end{lemma}

\vspace{6pt}
\begin{lemma}\cite{roger1994topics}\label{lemma hadamard probuct}
 For matrices $A\in \mathbb{R}^{m\times n}$ and  $B\in \mathbb{R}^{m\times n}$,
\begin{equation*}
\sigma(A\circ B)\leq \sigma(A)\sigma(B),
\end{equation*}
where $\sigma(\cdot)$ represents the maximum singular value of the corresponding matrix.
\end{lemma}

\vspace{6pt}
\begin{lemma}\cite{kadelburg2005inequalities}\label{lemma majorizaiton ineuqality} (Karamata's Inequality)
Let $f(\cdot)$ be a real-valued convex function defined on $I$, where $I$ is an interval of the real line, and
$x=(x_i)^n_{i=1}$ and $y=(y_i)^n_{i=1}$ are (finite) sequences of real numbers  in $I$. If   the sequence $x$ majorizes $y$, then
\begin{equation*}
\sum^n_{i=1}f(x_i)\geq \sum^n_{i=1}f(y_i).
\end{equation*}
\end{lemma}

%
%

\subsection{Problem Statement}

Consider a discrete-time linear stochastic system, which is measured by  a network of $N$ sensors, described by
\begin{equation}
\begin{aligned}
&x_{k+1} = Fx_{k}+\omega_{k},\\
& y_{i,k} = H_{i}x_{k}+\nu_{i,k}, ~~~i=1,2,...,N,
\end{aligned}
\end{equation}
where $k$ is the discrete time index,
$x_k\in \mathcal{R}^n$ is the system state vector,  $y_{i,k} \in \mathcal{R}^{m_i}$ is the measurement vector of sensor $i$, $F \in \mathcal{R}^{n\times n}$ is the state transition matrix,  $H_{i}\in \mathcal{R}^{m_i\times n}$ is the measurement matrix of sensor $i$, $\omega_{k}\in \mathcal{R}^n$  is the zero-mean Gaussian  process noise with the covariance $Q_{k}$, and $\nu_{i,k}$ is the zero-mean Gaussian measurement noise with the covariance $R_{i,k}$.  Suppose that $x_0$ is the initial state and $\Sigma_{i,0|0}$ is the initial estimation error covariance. The sequences $x_0$, $\{\omega_k\}^{\infty}_{k=0}$, and $\{\nu_{i,k}\}^{\infty,N}_{k=0,i=1}$ are mutually uncorrelated.

\vspace{6pt}
The  consensus-on-measurement  distributed filter (CMDF)\cite{olfati2007distributed,battistelli2014consensus} is a well-known  distributed algorithm to handle the distributed state estimation problem, presented in  Algorithm \ref{algorithm distributed}.
Generally, it is assumed that the covariance matrices $Q_{k}$ and $R_{i,k}$ are exactly known. However, in practical applications,
it is almost impossible for us to know the actual covariances $Q_{k}$ and $R_{i,k}$, and the filter will be ill-conditioned  with mismatched nominal covariances $Q^u_k$ and $R^u_{i,k}$.  When  the mismatched  covariances  are used, the performance of the  corresponding filter may be degraded. Hence, the objective of  this paper is to  study  the performance of the distributed  filter  in the presence of the mismatched noise covariances.

\vspace{6pt}
In this paper,   the   problem of CMDF  under the  mismatched noise covariances is formulated as follows:

\begin{enumerate}
\item  For three  performance evaluation indices  to evaluate the performance degradation
  based on CMDF presented in  Algorithm \ref{algorithm distributed} (which will be introduced in the next section), derive the difference expressions among these  indices.

\item  Establish  the one-step and the recursive  relations of the three performance evaluation indices under different noise covariance settings, and   evaluate how
     the consensus fusion affects  these relations.

\item  Analyze the convergence of  the  performance evaluation indices.  Find out  the bounds of   the performance of the asymptotic distributed filter by the noise covariance deviations and  these  performance evaluation indices.
\end{enumerate}

\section{different performance
evaluation indices}\label{section different estimation error covariance}
In this section,  three  performance
evaluation indices, including the standard and the nominal performance
evaluation indices, and the estimation error covariances, are constructed  to evaluate the performance of the  distributed filter under mismatched noise covariances, presented in   Algorithm \ref{algorithm distributed}, and the compact forms  and the   equivalent forms  are also given for further analysis.

\subsection{Standard Performance Evaluation Index}\label{sec 1 subsec standard}
The standard performance evaluation indices are defined as $\Sigma_{i,k|k-1}$ and $\Sigma_{i,k|k}$,  which are  the iterative matrices in Algorithm \ref{algorithm distributed} under the actual noise covariances $Q_{k}$ and $R_{i,k}$.

\vspace{6pt}
 Based on Algorithm \ref{algorithm distributed}, $\Sigma_{i,k|k}$ can be calculated  as
\begin{equation*}
\Sigma_{i,k|k} = \Big(\Sigma^{-1}_{i,k|k-1}+ N\sum^{N}_{j=1}l^{(L)}_{ij} (H^{(L)}_j)^TR^{-1}_{j,k}H^{(L)}_j\Big)^{-1}.
\end{equation*}

\vspace{6pt}
Define the modified measurement matrix, the modified measurement noise covariance matrix without the consensus terms, and
 the modified measurement noise covariance matrix with the consensus terms  as
\begin{equation*}
\tilde H^{(L)}_{i}=\Big(\text{sign}(l^{(L)}_{i1})H^T_{1},\ldots,
\text{sign}(l^{(L)}_{iN})H^T_{N}\Big)^T,~~~~~~~~\end{equation*}
\begin{equation}\label{eq std bar R}
\bar R^{(L)}_{i,k}=\text{diag}\Big(\text{sign}(l^{(L)}_{i1})R_{1,k},\ldots,\text{sign}(l^{(L)}_{iN})R_{N,k}\Big),
\end{equation}
and
\begin{equation}\label{eq std tilde R}
\tilde R^{(L)}_{i,k}=\text{diag}\Big(h^{(L)}_{i1}\text{sign}(l^{(L)}_{i1})R_{1,k},\ldots,h^{(L)}_{iN}\text{sign}(l^{(L)}_{iN})R_{N,k}\Big),
\end{equation}
respectively, where $h^{(L)}_{ij}=\frac{1}{Nl^{(L)}_{ij}}$ is set as $0$ if $Nl^{(L)}_{ij}=0$, and  $\text{sign}(x)$ is the sign function.
Then, $\Sigma_{i,k|k-1}$ and $\Sigma_{i,k|k}$ can be reformulated as
\begin{equation}\label{eq std sigma prior}
\Sigma_{i,k|k-1}= F\Sigma_{i,k-1|k-1}F^T+Q_{k-1},~~~~~~~~~~~~~~~~
\end{equation}
and
\begin{equation}\label{eq std sigma post1}
\Sigma_{i,k|k} = (\Sigma^{-1}_{i,k|k-1}+(\tilde H^{(L)}_i)^T (\tilde R^{(L)}_{i,k})^{-1}\tilde H^{(L)}_i)^{-1},~~
\end{equation}
respectively. In addition, $\Sigma_{i,k|k}$ has the equivalent form
\begin{equation}\label{eq std sigma 22}
\begin{aligned}
\Sigma_{i,k|k} =&\Sigma_{i,k|k-1}-\Sigma_{i,k|k-1}(\tilde H^{(L)}_i)^T(\tilde R^{(L)}_{i,k}\\
 &+\tilde H^{(L)}_i\Sigma_{i,k|k-1}(\tilde H^{(L)}_i)^T)^{-1}\tilde H^{(L)}_i\Sigma_{i,k|k-1}.
\end{aligned}
\end{equation}

\begin{algorithm}[]
\caption{CMDF}
\label{algorithm distributed}
\hspace*{0.02in} \textbf{Input:} $\hat{x}_{i, k-1|k-1}$, $\hat{\Sigma}_{i, k-1|k-1}$

\hspace*{0.02in} \textbf{Prediction:}
\begin{algorithmic}
\STATE $\hat x_{i,k|k-1}=F\hat x_{i,k-1|k-1}$
\STATE $\Sigma_{i,k|k-1}= F\Sigma_{i,k-1|k-1}F^T+Q_{k-1}$
\end{algorithmic}

\hspace*{0.02in} \textbf{Consensus fusion:}
\begin{algorithmic}
\STATE Set
\STATE ~~~~~~~$U^{(0)}_{i,k}=NH^T_{i}R^{-1}_{i,k}H_{i}$,~~~~~~~$V^{(0)}_{i,k}=NH^T_{i}R^{-1}_{i,k}y_{i,k}$
\STATE
\STATE For $m=1,2,...,L$
\STATE ~~~~~~~~~$U^{(m)}_{i,k}=\sum^{N}_{j=1}l_{ij}U^{(m-1)}_{j,k}$,~~~$V^{(m)}_{i,k}=\sum^N_{j=1}l_{ij}V^{(m-1)}_{j,k}$
\STATE
\end{algorithmic}

\hspace*{0.02in} \textbf{Correction:}
\begin{algorithmic}
\STATE $\Sigma_{i,k|k} = (\Sigma^{-1}_{i,k|k-1}+U^{(L)}_{i,k})^{-1}$
\STATE $\hat x_{i,k|k} = \Sigma_{i,k|k}(\Sigma^{-1}_{i,k|k-1}\hat x_{i,k|k-1}+V^{(L)}_{i,k})$
\end{algorithmic}

\hspace*{0.02in} \textbf{Output:}
\begin{algorithmic}
\STATE $\hat{x}_{i,k|k}$, $\Sigma_{i,k|k}$
\end{algorithmic}
\end{algorithm}

\subsection{Nominal Performance Evaluation Index}\label{sec subsec nominal}
The nominal performance evaluation indices are defined as
$\Sigma^f_{i,k|k-1}$ and $\Sigma^f_{i,k|k}$, and they  are the iterative matrices in Algorithm \ref{algorithm distributed} under
the  nominal process noise covariance  $Q^u_k$ and the nominal measurement noise covariances $R^u_{i,k}$, which are  practically used in the filter.

\vspace{6pt}
Define  $Q^u_k$ and $R^u_{i,k}$   as
\begin{equation*}
Q^u_{k} = Q_{k}+ \Delta Q_{k},~~~
\end{equation*}
and
\begin{equation*}
R^u_{i,k} = R_{i,k}+ \Delta R_{i,k},
\end{equation*}
respectively, where  $Q_{k}$ and $R_{i,k}$ are the actual process  noise covariance and measurement noise covariances, respectively, and $\Delta Q_{k}$ and $\Delta R_{i,k}$ are the corresponding deviations.
 Similarly to  Section \ref{sec 1 subsec standard}, define the modified nominal measurement noise covariance, the modified measurement noise covariance deviation, the modified measurement vector, and the modified measurement noise vector as
\begin{equation*}
\tilde R^{u(L)}_{i,k}=\text{diag}\Big(h^{(L)}_{i1}\text{sign}(l^{(L)}_{i1})R^u_{1,k},\ldots,
h^{(L)}_{iN}\text{sign}(l^{(L)}_{iN})R^u_{N,k}\Big),
\end{equation*}
\begin{equation*}
\Delta \bar R^{(L)}_{i,k}=\text{diag}\Big(\text{sign}(l^{(L)}_{i1})\Delta R_{1,k},\ldots,\text{sign}(l^{(L)}_{iN})\Delta R_{N,k}\Big),~~~~
\end{equation*}
\begin{equation*}
 \tilde y^{(L)}_{i,k} = \Big( \text{sign}(l^{(L)}_{i1})y^T_{1,k},\ldots, \text{sign}(l^{(L)}_{iN})y^T_{N,k}\Big)^T,
\end{equation*}
and
\begin{equation*}
 \tilde v^{(L)}_{i,k} = \Big( \text{sign}(l^{(L)}_{i1})v^T_{1,k},\ldots, \text{sign}(l^{(L)}_{iN})v^T_{N,k}\Big)^T,
\end{equation*}
respectively. Then, the superscript $f$ is  utilized to represent the nominal counterpart in the distributed filter, and
the following compact forms of the nominal distributed  filter  can be obtained as
\begin{equation*}
\hat x^f_{i,k|k-1}=F\hat x^f_{i,k-1|k-1},~~~~~~~~~~~~~~~~~~~~~
\end{equation*}
\begin{equation}\label{eq cal xf}
\begin{aligned}
\hat x^f_{i,k|k}
&= \Sigma^f_{i,k|k}(\Sigma^{f}_{i,k|k-1})^{-1}\hat x^f_{i,k|k-1}\\
&~~~~~+
\Sigma^f_{i,k|k}(\tilde H^{(L)}_{i})^T (\tilde R^{u(L)}_{i,k})^{-1}\tilde y^{(L)}_{i,k},
\end{aligned}
\end{equation}
\begin{equation}\label{eq cal sigma prior}
\Sigma^f_{i,k|k-1}= F\Sigma^f_{i,k-1|k-1}F^T+Q^u_{k-1},~~~~~
\end{equation}
and
\begin{equation}\label{eq cal sigmaf post1}
\Sigma^f_{i,k|k} = \big((\Sigma^{f}_{i,k|k-1})^{-1}+(\tilde H^{(L)}_i)^T (\tilde R^{u(L)}_{i,k})^{-1}\tilde H^{(L)}_i\big)^{-1}.
\end{equation}

For further analysis, the nominal gain matrix  $\tilde K^f_{i,k}$ and its relevant identities  are given   by performing   some algebraic manipulations
\begin{equation}\label{eq cal identi k}
\begin{aligned}
 \tilde K^f_{i,k} &= \Sigma^f_{i,k|k}(\tilde H^{(L)}_{i})^T (\tilde R^{u(L)}_{i,k})^{-1}\\
  &= \Sigma^f_{i,k|k-1}(\tilde H^{(L)}_{i})^T(\tilde R^{u(L)}_{i,k}+\tilde H^{(L)}_{i}\Sigma^f_{i,k|k-1}(\tilde H^{(L)}_{i})^T)^{-1},
\end{aligned}
\end{equation}
and
\begin{equation}\label{eq cal identi i-kh}
\begin{aligned}
I-\tilde K^f_{i,k}\tilde H^{(L)}_{i}&=
 \Sigma^f_{i,k|k}(\Sigma^f_{i,k|k-1})^{-1}\\
  &= I-\Sigma^f_{i,k|k-1}(\tilde H^{(L)}_{i})^T(\tilde R^{u(L)}_{i,k}\\
 &~~~~~~~+\tilde H^{(L)}_{i}\Sigma^f_{i,k|k-1}(\tilde H^{(L)}_{i})^T)^{-1}\tilde H^{(L)}_{i}.~~~~~~
\end{aligned}
\end{equation}


%

Then, by combining (\ref{eq cal identi k}) and (\ref{eq cal identi i-kh}), $\hat x^f_{i,k|k}$  and $\Sigma^f_{i,k|k}$ can be rewritten as
\begin{equation*}
\begin{aligned}
\hat x^f_{i,k|k} & = \hat x^f_{i,k|k-1}+   \tilde K^f_{i,k}(\tilde y^{(L)}_{i,k}-\tilde H^{(L)}_{i}\hat x^f_{i,k|k-1}),~~~~~
\end{aligned}
\end{equation*}
and
\begin{equation}\label{eq cal sigmaf2}
\begin{aligned}
\Sigma^f_{i,k|k} =& (I-\tilde K^f_{i,k}\tilde H^{(L)}_i)\Sigma^f_{i,k|k-1}(I-\tilde K^f_{i,k}\tilde H^{(L)}_i)^T\\
 &+ \tilde K^f_{i,k}\tilde R^{u(L)}_{i,k}(\tilde K^f_{i,k})^T,
\end{aligned}
\end{equation}
respectively.


\subsection{Estimation Error Covariance}
The estimation error covariances  are defined as $\Sigma^t_{i,k|k-1}$ and $\Sigma^t_{i,k|k}$, which are calculated by using  the state  estimate $\hat x^f_{i,k|k}$ under the nominal distributed  filter in Section \ref{sec subsec nominal}.

\vspace{6pt}
The superscript $t$  is used to  denote the actual counterpart.
Specifically,  the estimation errors and the estimation error covariance matrices are defined as
 $$e_{i,k|k-1} = x_k - \hat x^f_{i,k|k-1},~~~~~~~$$
 $$e_{i,k|k} = x_k - \hat x^f_{i,k|k},~~~~~~~~~~~~~$$ $$\Sigma^t_{i,k|k-1}=E\{e_{i,k|k-1}e^T_{i,k|k-1}\},$$ and $$\Sigma^t_{i,k|k}=E\{e_{i,k|k}e^T_{i,k|k}\},~~~~~~~~$$
 respectively. Recalling  the definition of  (\ref{eq std bar R}),  one has $\bar R^{(L)}_{i,k}=E\{\tilde\nu^{(L)}_{i,k}(\tilde\nu^{(L)}_{i,k})^T\}$.
  Then, according to (\ref{eq cal xf}),  the posterior  estimation error $e_{i,k|k}$ can be computed as
\begin{equation}\label{eq actual e pos}
\begin{aligned}
e_{i,k|k} & = \Sigma^f_{i,k|k}\Big((\Sigma^f_{i,k|k})^{-1}-\sum^N_{j=1}l^{(L)}_{ij}NH^T_{j}(R^{u}_{j,k})^{-1}H_{j}\Big)x_k\\
 &~~~-\Sigma^f_{i,k|k}(\Sigma^{f}_{i,k|k-1})^{-1}\hat x^f_{i,k|k-1}\\
 &~~~-\Sigma^f_{i,k|k}\sum^N_{j=1}l^{(L)}_{ij}NH^T_{j}(R^{u}_{j,k})^{-1}\nu_{j,k}\\
 &= \Sigma^f_{i,k|k}(\Sigma^f_{i,k|k-1})^{-1}e_{i,k|k-1}\\
 &~~~-\Sigma^f_{i,k|k} (\tilde H^{(L)}_{i})^T(\tilde R^{u(L)}_{i,k})^{-1}\tilde\nu^{(L)}_{i,k}.
\end{aligned}
\end{equation}
Similarly, the prior  estimation error $e_{i,k|k-1}$ can be  given as
\begin{equation}\label{eq actual e pir}
\begin{aligned}
e_{i,k|k-1}= Fe_{i,k-1|k-1}+\omega_{k-1}.
\end{aligned}
\end{equation}

\vspace{6pt}
Based on  (\ref{eq actual e pos}) and (\ref{eq actual e pir}), $\Sigma^t_{i,k|k}$ and $\Sigma^t_{i,k|k-1}$ can be computed as
\begin{equation}\label{eq actual sigma1}
\begin{aligned}
 \Sigma^t_{i,k|k} &=  \Sigma^f_{i,k|k}(\Sigma^f_{i,k|k-1})^{-1}\Sigma^t_{i,k|k-1}(\Sigma^f_{i,k|k-1})^{-1}\Sigma^f_{i,k|k}\\
&~~~+\Sigma^f_{i,k|k} (\tilde H^{(L)}_{i})^T(\tilde R^{u(L)}_{i,k})^{-1} \bar R^{(L)}_{i,k}\\
&~~~\times(\tilde R^{u(L)}_{i,k})^{-1}\tilde H^{(L)}_{i}\Sigma^f_{i,k|k},
\end{aligned}
\end{equation}
and
\begin{equation}\label{eq actual sigma prior}
\begin{aligned}
 \Sigma^t_{i,k|k-1} = F\Sigma^t_{i,k-1|k-1}F^T+Q_{k-1},
\end{aligned}
\end{equation}
respectively. By combining (\ref{eq cal identi k}), (\ref{eq cal identi i-kh}) with  (\ref{eq actual sigma1}), $\Sigma^t_{i,k|k}$ has the following equivalent form
\begin{equation}\label{eq actual sigma2}
\begin{aligned}
 \Sigma^t_{i,k|k} =&
 (I-\tilde K^f_{i,k}\tilde H^{(L)}_i)\Sigma^t_{i,k|k-1}(I-\tilde K^f_{i,k}\tilde H^{(L)}_i)^T\\
 &+ \tilde K^f_{i,k}\bar R^{(L)}_{i,k}(\tilde K^f_{i,k})^T.
\end{aligned}
\end{equation}

\section{Relations among the three  performance evaluation indices }\label{sec analyze difference }

This section discusses the relations among  $\Sigma^t_{i,k|k}$,  $\Sigma^f_{i,k|k}$, and $\Sigma_{i,k|k}$ of CMDF  under mismatched noise covariances. First, the difference expressions are derived based on the above section. Then, the relation between $\Sigma^t_{i,k|k}$ and $\Sigma_{i,k|k}$, which is representative and almost deterministic, is discussed, and the effect of the consensus terms is also analyzed. Finally, other relations are discussed.

\vspace{6pt}
To show the effect of the mismatched noise covariances in detail,  the following assumptions  are made.

\vspace{6pt}
\begin{assumption}\label{ass k-1 equal}
Three   performance evaluation indices $\Sigma^t_{i,k-1|k-1}$, $\Sigma^f_{i,k-1|k-1}$, and $\Sigma_{i,k-1|k-1}$ are the same  at time step $k-1$, i.e.,
$\Sigma^t_{i,k-1|k-1} = \Sigma^f_{i,k-1|k-1}= \Sigma_{i,k-1|k-1}$.   In addition, $R_{i,k}$, $R^u_{i,k}$, $Q_{k-1}$, and $Q^u_{k-1}$ are positive definite.
\end{assumption}

\vspace{6pt}
\begin{assumption}\label{ass communication topology}
The communication topology is  strongly connected.
\end{assumption}

\vspace{6pt}
First, the relations among the different  predicted  performance evaluation indices  can be directly  obtained  under the  deviation $\Delta Q_k$.
If  $\Delta Q_{k-1}\geq 0$ and Assumption \ref{ass k-1 equal} holds, then $\Sigma^t_{i,k|k-1} = \Sigma_{i,k|k-1}\leq \Sigma^f_{i,k|k-1}$ by combining (\ref{eq std sigma prior}), (\ref{eq cal sigma prior}), and (\ref{eq actual sigma prior}).  Conversely, if $\Delta Q_{k-1}\leq 0$, then $\Sigma^t_{i,k|k-1} = \Sigma_{i,k|k-1}\geq \Sigma^f_{i,k|k-1}$.

\vspace{6pt}
Next, the relations among the different posterior  performance evaluation indices, i.e., $\Sigma^t_{i,k|k}$,  $\Sigma^f_{i,k|k}$, and $\Sigma_{i,k|k}$,  are considered  under the deviations $\Delta Q_{k-1}$ and $\Delta R_{i,k}$.
To better show their relations,
define  $\Phi^f_{i,k}$,  $\Phi_{i,k}$, and $\Phi^t_{i,k}$ as follows
\begin{equation}\label{eq phif}
\begin{aligned}
\Phi^f_{i,k} &= (\tilde H^{(L)}_{i})^T(\tilde R^{u(L)}_{i,k})^{-1}\tilde H^{(L)}_{i}\\
&=\sum^{N}_{j=1} Nl^{(L)}_{ij}H^T_{j}(R^{u}_{j,k})^{-1}H_{j},
\end{aligned}
\end{equation}
\begin{equation}\label{eq phis}
\begin{aligned}
\Phi_{i,k} &= (\tilde H^{(L)}_{i})^T(\tilde R^{(L)}_{i,k})^{-1}\tilde H^{(L)}_{i}\\
&=\sum^{N}_{j=1} Nl^{(L)}_{ij}H^T_{j}(R_{j,k})^{-1}H_{j},
\end{aligned}
\end{equation}
and
\begin{equation}\label{eq phit}
\begin{aligned}
\Phi^t_{i,k} &= (\tilde H^{(L)}_{i})^T(\tilde R^{u(L)}_{i,k})^{-1} \bar R^{(L)}_{i,k}(\tilde R^{u(L)}_{i,k})^{-1}\tilde H^{(L)}_{i}\\
&=\sum^{N}_{j=1} N^2(l^{(L)}_{ij})^2H^T_{j}(R^{u}_{j,k})^{-1}R_{j,k}(R^{u}_{j,k})^{-1}H_{j},
\end{aligned}
\end{equation}
respectively.

\subsection{The Difference Expressions}


The difference expressions among $\Sigma^t_{i,k|k}$, $\Sigma_{i,k|k}$,  and $\Sigma^f_{i,k|k}$
are given in this subsection.  To simplify the derivation, it is assumed that $\Delta R_{i,k}$ is invertible.

\vspace{6pt}
\begin{proposition}\label{lemma sigmat-sigma}
Under  Assumption \ref{ass k-1 equal}, the difference between $\Sigma^t_{i,k|k}$ and $\Sigma_{i,k|k}$ is  $$\Sigma^t_{i,k|k} - \Sigma_{i,k|k} = \tilde  \Sigma^t_{i,k|k} - \Sigma_{i,k|k}+ R^{ts}_{i,k|k},$$
where $\tilde  \Sigma^t_{i,k|k} - \Sigma_{i,k|k}\geq 0$,
\begin{equation}\label{eq lemma sigmat-sigma 1}
\begin{aligned}
\tilde  \Sigma^t_{i,k|k} =&
 (I-\tilde K^f_{i,k}\tilde H^{(L)}_i)\Sigma^t_{i,k|k-1}(I-\tilde K^f_{i,k}\tilde H^{(L)}_i)^T\\
 &+ \tilde K^f_{i,k}\tilde R^{(L)}_{i,k}(\tilde K^f_{i,k})^T,
\end{aligned}
\end{equation}
\begin{equation}\label{eq lemma sigmat-sigma 2}
\begin{aligned}
 R^{ts}_{i,k|k}&= \Sigma^f_{i,k|k-1}
 \bar C_{i,k}\Phi^{ts}_{i,k}\bar C^T_{i,k}(\Sigma^f_{i,k|k-1})^T,~~~~~
\end{aligned}
\end{equation}
\begin{equation}\label{eq lemma sigmat-sigma 3}
\begin{aligned}
\bar C_{i,k}=& I-(\tilde H^{(L)}_i)^T(\tilde R^{u(L)}_{i,k})^{-1}\tilde H^{(L)}_i((\Sigma^f_{i,k|k-1})^{-1}\\
&+(\tilde H^{(L)}_i)^T(\tilde R^{u(L)}_{i,k})^{-1}\tilde H^{(L)}_i
)^{-1},
\end{aligned}
\end{equation}
and
\begin{equation}\label{eq lemma sigmat-sigma 4}
\begin{aligned}
\Phi^{ts}_{i,k}&= \sum^{N}_{j=1} \big((Nl^{(L)}_{ij})^2-Nl^{(L)}_{ij}\big)H^T_{j}(R^u_{j,k})^{-1}R_{j,k}(R^u_{j,k})^{-1}H_{j}.~~~~~~~~~~~~~
\end{aligned}
\end{equation}
\end{proposition}

The proof of Proposition \ref{lemma sigmat-sigma} is given in  Appendix    \ref{appdex lemma sigmat-sigma}.

\vspace{6pt}
\begin{proposition}\label{lemma sigmat-sigmaf}
Under  Assumption \ref{ass k-1 equal},
the difference between $\Sigma^t_{i,k|k}$ and $\Sigma^f_{i,k|k}$ is  $$\Sigma^t_{i,k|k} - \Sigma^f_{i,k|k} =\Sigma^f_{i,k|k}(\Phi^t_{i,k} - \Phi^f_{i,k})\Sigma^f_{i,k|k} + \Psi^{tf}_{i,k},$$
 where
\begin{equation}\label{eq lemma sigmat -sigmaf 1}
\begin{aligned}
\Phi^t_{i,k}-\Phi^f_{i,k}
&= \sum^{N}_{j=1} Nl^{(L)}_{ij}H^T_{j}(R^{u}_{j,k})^{-1}
((Nl^{(L)}_{ij}-1)I\\
&~~~~-Nl^{(L)}_{ij}\Delta R_{j,k}(R^{u}_{j,k})^{-1})H_{j},
\end{aligned}
\end{equation}
and
\begin{equation}\label{eq lemma sigmat -sigmaf 1-1-1}
\begin{aligned}
\Psi^{tf}_{i,k} = -(I-\tilde K^f_{i,k}\tilde H^{(L)}_i)\Delta Q_{k-1}(I-\tilde K^f_{i,k}\tilde H^{(L)}_i).
\end{aligned}
\end{equation}

\end{proposition}

The proof of Proposition \ref{lemma sigmat-sigmaf} is given in  Appendix \ref{appdex lemma sigmat-sigmaf}.

\vspace{6pt}
\begin{proposition}\label{lemma sigmaf-sigma}
Under  Assumption \ref{ass k-1 equal},
the difference between $\Sigma^f_{i,k|k}$ and $\Sigma_{i,k|k}$ is $$(\Sigma^f_{i,k|k})^{-1}- (\Sigma_{i,k|k})^{-1} = \Phi^f_{i,k}-\Phi_{i,k}+\Psi^{fs}_{i,k},$$
where
\begin{equation}\label{eq lamma sigmaf-sigmas 1}
\begin{aligned}
\Phi^f_{i,k}-\Phi_{i,k} &= \sum^{N}_{j=1} Nl^{(L)}_{ij}H^T_{j}\big(
-R^{-1}_{j,k}\Delta R_{j,k}(R^{u}_{j,k})^{-1}
\big)H_{j},
\end{aligned}
\end{equation}
and
\begin{equation}\label{eq lemma simgf-sigmas psi}
\begin{aligned}
\Psi^{fs}_{i,k} =  (\Sigma^f_{i,k|k-1})^{-1}- (\Sigma_{i,k|k-1})^{-1}.
\end{aligned}
\end{equation}
\end{proposition}

The proof of Proposition \ref{lemma sigmaf-sigma} is given in  Appendix \ref{appdex lemma sigmaf-sigma}.

\vspace{6pt}
\begin{remark}
To simplify the derivation, the assumption that  $\Delta R_{i,k}$ is invertible  has been made. However, when $\Delta R_{i,k}$  has zero eigenvalues,  the corresponding expressions can be also derived.   Based on Assumption \ref{ass k-1 equal},
 $\Delta R_{i,k}$ is real symmetric, and   $\Delta R_{i,k}$  can be orthogonal diagonalization. Then,   similar expressions can be derived by eliminating the corresponding rows of  zero eigenvalues of $\Delta R_{i,k}$.
 Hence, the relations can be directly  compared
 by $\Phi^f_{i,k}$, $\Phi^t_{i,k}$, and $\Phi_{i,k}$, and this assumption has no impact on  the results.
\end{remark}

\subsection{Relation Analysis Between $\Sigma^t_{i,k|k}$ and $\Sigma_{i,k|k}$ }

This subsection  first studies the relation between   $\Sigma^t_{i,k|k}$ and $\Sigma_{i,k|k}$, since this relation
is representative and almost deterministic.  Then, the effect of the consensus terms in this relation is  analyzed. It is worth mentioning that  the analysis of the effect of the consensus terms can be also  applied in other relations.

\vspace{6pt}
Based on Proposition  \ref{lemma sigmat-sigma},  the relation between  $\Sigma^t_{i,k|k}$  and $\Sigma_{i,k|k}$ depends on $\tilde  \Sigma^t_{i,k|k} - \Sigma_{i,k|k}$ and $\Phi^{ts}_{i,k}$, and it has been proven that $\tilde  \Sigma^t_{i,k|k} - \Sigma_{i,k|k}\geq 0$. Next,
the properties of  $\Phi^{ts}_{i,k}$
are studied, and it is obvious that  $(Nl^{(L)}_{ij})^2-Nl^{(L)}_{ij}$ plays  a key role in determining whether $\Phi^{ts}_{i,k}$ is positive definite or negative definite.

\vspace{6pt}
\begin{proposition}\label{lemma single sigmat-sigma 0}
  Under Assumption \ref{ass k-1 equal}  and considering the case that $N=1$, it holds    $\Sigma^t_{1,k|k}\geq \Sigma_{1,k|k}$.
\end{proposition}

The proof of Proposition \ref{lemma single sigmat-sigma 0}  is given in  Appendix \ref{appdex lemma single sigmat-sigma 0}.

\vspace{6pt}
For the state estimation problem involving a single sensor, $\Sigma^t_{1,k|k}\geq \Sigma_{1,k|k}$ holds under any  mismatched noise covariance scenario.
However, for the distributed filtering based on sensor networks, the relation between
 $\Sigma^t_{i,k|k}$  and $\Sigma_{i,k|k}$ is  affected   by the consensus  fusion. The following example shows this feature.

\vspace{6pt}
\begin{example}\label{example 1}
 Consider   a network of three sensors measuring the state of a scalar system, whose communication topology is depicted in Fig. \ref{example 1}. The parameters are set as $F = 1$, $H_{i} = 1, ~~i\in \{1, 2, 3\}$,
$R_{1, k}= R^u_{1, k}=R_{2, k}=R^u_{2, k}=1$, $R_{3, k} = 0.1$, $R^u_{3, k} = 0.11$, $Q_{k-1} =1$, $Q^u_k =2$, and $\Sigma^t_{i,k-1|k-1} = \Sigma^f_{i,k-1|k-1}= \Sigma_{i,k-1|k-1} =4$.  The Metropolis weights are adopted to construct  the adjacent matrix $\mathcal{L}$, and $L$ represents the fusion step.

\begin{figure}[!htb]
\centering
{\includegraphics[width=1.5in]{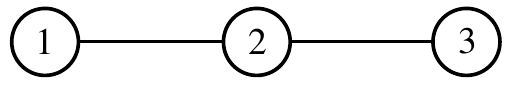}}
\caption{The diagram of the communication topology in Example \ref{example 1}.}
\label{fig communication_topology example}
\end{figure}

Then, the estimation error  covariances after one-step iteration can be computed.  When $L=2$,  $\Sigma^t_{1,k|k}=0.1406$, $\Sigma^t_{2,k|k}=0.0821$, $\Sigma^t_{3,k|k}=0.0873$, $\Sigma_{1,k|k}=0.1613$, $\Sigma_{2,k|k}=0.0820$, and $\Sigma_{3,k|k}=0.0549$.
It can be seen that $\Sigma^t_{1,k|k}<\Sigma_{1,k|k}$, $\Sigma^t_{2,k|k}>\Sigma_{2,k|k}$, and $\Sigma^t_{3,k|k}>\Sigma_{3,k|k}$.
\end{example}

\vspace{6pt}
Based on Example \ref{example 1}, it is obvious that
 Proposition \ref{lemma single sigmat-sigma 0} is not applicable for sensor netwroks. It is reasonable that the estimation error covariance is larger than the standard performance evaluation index under mismatched noise covariances, like Proposition \ref{lemma single sigmat-sigma 0}. However, it is shown  that  the consensus-based information fusion  affects    the relations. Next, the effect of the  consensus terms  is analyzed.

\vspace{6pt}
\begin{proposition}\label{lemma barl approach 0}
Define $\bar l^{(m)}_{ij} = (Nl^{(m)}_{ij})^2-Nl^{(m)}_{ij}$ and
$[\mathcal{\bar L}^{m}]_{ij} = \bar l^{(m)}_{ij}$.
If Assumption
\ref{ass communication topology} holds and $\mathcal{L}$ is symmetric and doubly stochastic with the positive diagonal elements, then the matrix  $\mathcal{\bar L}^m$ converges exponentially to $0$ as $m$ tends to infinity.
\end{proposition}

The proof of Proposition \ref{lemma barl approach 0} is given in  Appendix
\ref{appdex lemma barl approach 0}.

%
%

\vspace{6pt}
\begin{theorem}\label{lemma katamata inequality}
If Assumption
\ref{ass communication topology} holds and $\mathcal{L}$ is doubly stochastic with  positive diagonal elements, then
for a number $\gamma = 0$ or 1 and
any step $m \geq d$, the following results hold
\begin{enumerate}
\item For  $l^{(m)}_{ij}$ and $l^{(d)}_{ij}$,
\begin{equation}\label{eq lemma katamata ine 1}
\begin{aligned}
(1-\gamma) N&\leq\sum^N_{j=1}((Nl^{(m)}_{ij})^2-\gamma Nl^{(m)}_{ij})\\
&\leq \sum^N_{j=1}((Nl^{(d)}_{ij})^2-\gamma Nl^{(d)}_{ij}).
\end{aligned}
\end{equation}

\item For any element $l^{(m)}_{ij}$,  every element of the matrix $\mathcal{L}^m$ has  the following bounds
\begin{equation*}
\begin{aligned}
 \underline l  \leq [\mathcal{L}^m]_{ij} \leq \bar l,
\end{aligned}
\end{equation*}
where  $\underline l$ and $\bar l$ are defined as   $\underline l = \text{min}_{1\leq p,q\leq N} [\mathcal{L}^d]_{p,q}$ and $\bar l =\text{max}_{1\leq p,q\leq N} [\mathcal{L}^d]_{p,q}$, respectively.
Moreover, for $\bar l$ and $\underline l$, it holds that
\begin{equation*}
\begin{aligned}
 \frac{1}{N}\leq \bar l \leq 1, ~~ 0\leq \underline l\leq \frac{1}{N}.
\end{aligned}
\end{equation*}
In addition, $\bar l\to \frac{1}{N}$ and $\underline l \to \frac{1}{N}$ as $d\to \infty$.

\item

For $\bar l^{(m)}_{ij} = (Nl^{(m)}_{ij})^2-\gamma Nl^{(m)}_{ij}$ and
$[\mathcal{\bar L}^{m}]_{ij} = \bar l^{(m)}_{ij}$,
every element of the matrix $\mathcal{\bar L}^m$ has  the following bounds
\begin{equation}\label{eq lemma katamata ine 3}
\begin{aligned}
 \text{min}\{-\frac{\gamma}{4}, (N\underline l)^2-\gamma N\underline l \}\leq [\mathcal{\bar L}^m]_{ij} \leq (N\bar l)^2-\gamma N\bar l.
\end{aligned}
\end{equation}
Consequently,   $\bar l^{(m)}_{ij}\to (1- \gamma) $ as  $d\to \infty$.


\end{enumerate}
\end{theorem}

The proof of Theorem \ref{lemma katamata inequality} is given in  Appendix \ref{appdex lemma katamata inequality}.

\vspace{6pt}
\begin{remark}
 Proposition \ref{lemma barl approach 0} and Theorem \ref{lemma katamata inequality}  provide a perspective on how the consensus  terms work on  the relations among these performance evaluation indices, and the properties of  the sums, the bounds, and the convergence of the consensus terms with the increasing fusion step are discussed. By using the number $\gamma$, the properties of
 $(Nl^{(m)}_{ij})^2-Nl^{(m)}_{ij}$ and
 $(Nl^{(m)}_{ij})^2$,
  which can be founded in Propositions  \ref{lemma sigmat-sigma}, \ref{lemma sigmat-sigmaf} and Proposition \ref{lemma sigmat-sigmaf}, respectively,
  are  integrated into  Theorem \ref{lemma katamata inequality}.
 The theories  of the Hadamard product and the majorization are utilized to obtain these properties.
\end{remark}

\vspace{6pt}
\begin{lemma}\label{lemma phito0}
Under  Assumptions
 \ref{ass k-1 equal} and \ref{ass communication topology},  one has $\Phi^{ts}_{i,k} \to 0$ and $R^{ts}_{i,k|k}\to 0$ as $L\to \infty$.
\end{lemma}

The proof of Lemma \ref{lemma phito0} is given in  Appendix \ref{appdex lemma phito0}.

\vspace{6pt}
\begin{theorem}\label{theorem Sigmat-sigma}
Under Assumption \ref{ass k-1 equal} and \ref{ass communication topology}, it follows
\begin{enumerate}
\item If $\Phi^{ts}_{i,k}\geq 0$, it holds $\Sigma^t_{i,k|k}\geq \Sigma_{i,k|k}$.
\item There exists a non-positive sequence  $\varepsilon^{(L)}_k$ such that $\Sigma^t_{i,k|k}-\Sigma_{i,k|k}\geq \varepsilon^{(L)}_k I$ always  holds.  Moreover,  as $L \to \infty$,  $\varepsilon^{(L)}_k\to 0$.

\end{enumerate}
\end{theorem}

The proof of Theorem \ref{theorem Sigmat-sigma} is given in  Appendix \ref{appdex theorem Sigmat-sigma}.

\vspace{6pt}
\begin{corollary}\label{corollary sigmt sigma}
 Under Assumptions \ref{ass k-1 equal} and \ref{ass communication topology}  and considering the case that
 the terms $H^T_{j}(R^u_{j,k})^{-1}R_{j,k}(R^u_{j,k})^{-1}H_{j}$ of
 all sensors are the same, for any step $m \geq d$, it holds
$0\leq \Phi^{ts(m)}_{i,k} \leq \Phi^{ts(d)}_{i,k}$ and $\Sigma^t_{i,k|k}\geq \Sigma_{i,k|k}$.
\end{corollary}

The proof of Corollary \ref{corollary sigmt sigma} is given in  Appendix \ref{appdex corollary sigmt sigma}.

\vspace{6pt}
\begin{remark}\label{remark sigmat sigma 1}
 Proposition \ref{lemma single sigmat-sigma 0} and  Example \ref{example 1} show that the consensus terms  make $\Sigma^t_{i,k|k}<\Sigma_{i,k|k}$ possible for  sensor networks.
 However, Theorem  \ref{lemma katamata inequality}-1 and Corollary \ref{corollary sigmt sigma} imply that  $\Phi^{ts}_{i,k}$ has a tendency to be greater than 0.  Proposition  \ref{lemma barl approach 0}, Theorem \ref{lemma katamata inequality}, and Lemma \ref{lemma phito0} illustrate that as the fusion step $L$ increases,  the effect of the consensus term will gradually decrease.
\end{remark}

\subsection{Relation Analysis Among $\Sigma^t_{i,k|k}$, $\Sigma^f_{i,k|k}$ and $\Sigma_{i,k|k}$ }

This subsection focuses on the relations among   $\Sigma^t_{i,k|k}$, $\Sigma^f_{i,k|k}$ and $\Sigma_{i,k|k}$. 
First, the relations among $\Phi^t_{i,k}$, $\Phi^f_{i,k}$, $\Phi_{i,k}$, and $\Phi^{ts}_{i,k}$ are studied  for further analysis.

%
%
%
%
%
%
%
%

%
%

\vspace{6pt}
\begin{lemma}\label{lemma phits phit-phif 1}
 For any consensus fusion step $L$, it holds that
$\Phi^{ts}_{i,k}- (\Phi^t_{i,k}-\Phi^f_{i,k}) =
\bar \Phi^{tf}_{i,k}
$, where $\bar \Phi^{tf}_{i,k} =\sum^{N}_{j=1} Nl^{(L)}_{ij}H^T_{j}(R^u_{j,k})^{-1} \Delta R_{j,k}(R^u_{j,k})^{-1}H_{j}$.  As $L\to \infty$, $\lim_{L\to \infty} \Phi^{ts}_{i,k}=0$ and $\lim_{L\to \infty} (\Phi^t_{i,k}-\Phi^f_{i,k})=-\bar \Phi^{tf}_{i,k}$.
If $\Delta R_{i,k}=0$ for all $i \in \mathcal{V}$, then $\Phi^{ts}_{i,k}= \Phi^t_{i,k}-\Phi^f_{i,k}$.
\end{lemma}

The proof of Lemma \ref{lemma phits phit-phif 1} is given in  Appendix \ref{appdex lemma phits phit-phif 1}.


\vspace{6pt}
\begin{lemma}\label{lemma phif=phi}
 If $\Delta R_{i,k}=0$ for all $i \in \mathcal{N}$,  $\Phi^f_{i,k} = \Phi_{i,k}$. In addition,   $\Phi^t_{i,k} \to \Phi_{i,k}$ as $L\to \infty$.
\end{lemma}

\vspace{6pt}
\begin{proof}
The proof is straightforward based on  (\ref{eq phit}), (\ref{eq phif}),  (\ref{eq phis}) and  Proposition \ref{lemma barl approach 0}.
\end{proof}


Lemma \ref{lemma phits phit-phif 1}-\ref{lemma phif=phi}  show the relations among $\Phi^t_{i,k}$, $\Phi^f_{i,k}$, $\Phi_{i,k}$, and $\Phi^{ts}_{i,k}$, which provide powerful tools to investigate the relations  among $\Sigma^t_{i,k|k}$, $\Sigma^f_{i,k|k}$ and $\Sigma_{i,k|k}$. Next, under different noise  covariance deviation settings, these relations are discussed.

\vspace{6pt}
\begin{theorem}\label{theorem phi q=0}
   Under Assumptions \ref{ass k-1 equal} and \ref{ass communication topology} and  considering the case that   $\Delta Q_{k-1} = 0$,  it follows

\begin{enumerate}
\item  If  $\Phi_{i,k} \leq \Phi^f_{i,k}\leq \Phi^t_{i,k}$   and  $\Phi^{ts}_{i,k}\geq 0$, then  $\Sigma^f_{i,k|k} \leq \Sigma_{i,k|k} \leq \Sigma^t_{i,k|k}$.

\item If  $\Phi^f_{i,k} \leq \Phi_{i,k}$, $ \Phi^f_{i,k} \leq \Phi^t_{i,k}$   and  $\Phi^{ts}_{i,k}\geq 0$,  then $\Sigma_{i,k|k} \leq \Sigma^f_{i,k|k} \leq \Sigma^t_{i,k|k}$.

\item If  $\Phi^t_{i,k} \leq \Phi^f_{i,k} \leq \Phi_{i,k}$   and  $\Phi^{ts}_{i,k}\geq 0$, then $\Sigma_{i,k|k} \leq \Sigma^t_{i,k|k} \leq \Sigma^f_{i,k|k}$.
\end{enumerate}

\end{theorem}

The proof of Theorem \ref{theorem phi q=0} is given in  Appendix \ref{appdex theorem phi q=0}.

\vspace{6pt}

\begin{theorem}\label{theorem deltar q=0}
  Under Assumptions \ref{ass k-1 equal} and \ref{ass communication topology} and considering the case  that  $\Delta Q_{k-1} = 0$,  it follows
\begin{enumerate}
\item  If there exists at least one $\Delta R_{i,k}>0$ and $\Delta R_{i,k}\geq 0,~\forall i\in \mathcal{V}$, then
   $\Sigma_{i,k|k}<\Sigma^t_{i,k|k}<\Sigma^f_{i,k|k}$ as $L\to \infty$.

\item  If there exists at least one $\Delta R_{i,k}<0$ and $\Delta R_{i,k}\leq 0,~\forall i\in \mathcal{V}$, then
$\Sigma^f_{i,k|k}<\Sigma_{i,k|k}<\Sigma^t_{i,k|k}$ as $L\to \infty$.

\item  If  $\Delta R_{i,k}=0,~\forall i\in \mathcal{V}$, then
$\Sigma^f_{i,k|k}=\Sigma_{i,k|k}$. Moreover,
$\Sigma^t_{i,k|k}$ converges to  $\Sigma^f_{i,k|k}$ and $\Sigma_{i,k|k}$  as  $L\to \infty$.
\end{enumerate}
\end{theorem}

The proof of Theorem \ref{theorem deltar q=0} is given in  Appendix \ref{appdex theorem deltar q=0}.

\vspace{6pt}
\begin{theorem}\label{theorem r=0}
   Under Assumptions \ref{ass k-1 equal} and \ref{ass communication topology} and considering the case  that  $\Delta R_{i,k} = 0$,  it follows
\begin{enumerate}
\item  If $\Delta Q_{k-1}> 0$, then  $\Sigma_{i,k|k} < \Sigma^f_{i,k|k}$. Moreover, it holds $\Sigma_{i,k|k} <\Sigma^t_{i,k|k} < \Sigma^f_{i,k|k}$ as $L\to \infty$.

\item  If  $\Delta Q_{k-1} < 0$, then  $\Sigma_{i,k|k} > \Sigma^f_{i,k|k}$. Moreover, it holds $\Sigma^f_{i,k|k} \leq \Sigma_{i,k|k} \leq \Sigma^t_{i,k|k}$ as $L\to \infty$.
\end{enumerate}
\end{theorem}

The proof of Theorem \ref{theorem r=0}  is given in  Appendix \ref{appdex theorem r=0}.

\vspace{6pt}
\begin{theorem}\label{theorem q r}
  Under Assumptions \ref{ass k-1 equal} and \ref{ass communication topology}, it follows

\begin{enumerate}
\item  If  $\Phi_{i,k} \leq \Phi^f_{i,k}\leq \Phi^t_{i,k}$, $\Phi^{ts}_{i,k}\geq 0$, and $\Delta Q_{k-1}< 0$, then  $\Sigma^f_{i,k|k} \leq \Sigma_{i,k|k} \leq \Sigma^t_{i,k|k}$.

\item If  $\Phi^t_{i,k} \leq \Phi^f_{i,k} \leq \Phi_{i,k}$, $\Phi^{ts}_{i,k}\geq 0$, and $\Delta Q_{k-1}> 0$, then $\Sigma_{i,k|k} \leq \Sigma^t_{i,k|k} \leq \Sigma^f_{i,k|k}$.
\end{enumerate}
\end{theorem}

\vspace{6pt}
\begin{proof}
The proof is straightforward by combining Theorems \ref{theorem phi q=0} and \ref{theorem r=0}.
\end{proof}

\vspace{6pt}
The above theorems show the one-step relations among three
performance evaluation indices. Next,  the recursive relations with the increasing time step
are considered.

\vspace{6pt}
\begin{theorem}\label{theorem recurse}
   Under Assumptions \ref{ass k-1 equal} and \ref{ass communication topology}, for any time step $m$ satisfying  $m\geq k$,  it follows
\begin{enumerate}
\item  If  $\Phi_{i,j} \leq \Phi^f_{i,j}\leq \Phi^t_{i,j}$, $\Phi^{ts}_{i,j}\geq 0$, and $\Delta Q_{j-1}< 0$ hold for all $j\in [k,m]$, then  $\Sigma^f_{i,m|m} \leq \Sigma_{i,m|m} \leq \Sigma^t_{i,m|m}$.

\item If  $\Phi^t_{i,j} \leq \Phi^f_{i,j} \leq \Phi_{i,j}$, $\Phi^{ts}_{i,j}\geq 0$, and $\Delta Q_{j-1}> 0$ hold for all $j\in [k,m]$, then $\Sigma_{i,m|m} \leq \Sigma^t_{i,m|m} \leq \Sigma^f_{i,m|m}$.
\end{enumerate}
\end{theorem}

The proof of Theorem \ref{theorem recurse} is given in  Appendix \ref{appdex theorem recurse}.

\vspace{6pt}
\begin{remark}
Under  the different noise covariance  settings,
more recursive relations can be obtained   based on   Theorems \ref{theorem phi q=0}-\ref{theorem r=0}  by using the  similar techniques in Theorem  \ref{theorem recurse}.    For simplicity,  only one theorem is presented.   From the one-step results to the multi-step results,
 recursive relations are more general and meaningful.
It is worth mentioning that the changes of   these relations with the increasing fusion step can be  founded by
 utilizing Proposition \ref{lemma barl approach 0} and Theorem \ref{lemma katamata inequality}.

\end{remark}

\vspace{6pt}
\begin{remark}
 In  Theorems \ref{theorem phi q=0}, \ref{theorem q r} and \ref{theorem recurse},  it is assumed that $\Phi^{ts}_{i,k}\geq 0$ and $\Sigma_{i,k|k} \leq \Sigma^t_{i,k|k}$, and there may exist some conservatism. When $\Phi^{ts}_{i,k}< 0$, the corresponding  conclusions can be only drawn by directly using the relations between $\Sigma^t_{i,k|k}$ and $\Sigma_{i,k|k}$, since    $\Phi^{ts}_{i,k}$ are affected by
 the positive semi-define term $\tilde  \Sigma^t_{i,k|k} - \Sigma_{i,k|k}$ in Proposition \ref{lemma sigmat-sigma}.
  However,   Lemma \ref{lemma phito0}
 shows that  $\Phi^{ts}_{i,k}$ converges to $0$ as $L$ tends to infinity,  and Theorem \ref{lemma katamata inequality}  implies that  $\Phi^{ts}_{i,k}$ has a tendency to be greater than 0. In addition,  the term $\tilde  \Sigma^t_{i,k|k} - \Sigma_{i,k|k}$, which is greater than 0, reduces the impact of $\Phi^{ts}_{i,k}$. Hence,  this assumption is acceptable and reasonable.
\end{remark}

\vspace{6pt}
\begin{remark}
Based on the above results, more criterions can be used to evaluate the relations among three  performance evaluation indices, such as  the trace and the Frobenius norm.
The different criterions focus on the different aspects of the performance evaluation matrices, and provide different perspective on these relations.
\end{remark}

\section{Convergence Analysis}\label{sec convergence analysis}

This section  studies the convergence of the nominal performance evaluation index  and the estimation error covariance, and derives the bounds of the performances of  the asymptotic distributed filter  related to $\Delta Q$ and $\Delta R_i$.

\vspace{6pt}
For the convergence problem, it is assumed that  $Q_{k}$, $Q^u_{k}$, $R_{i,k}$, $R^u_{i,k}$  are constant,
 and the subscript $k$ is dropped  to represent the constant matrix, i.e.,  $Q$, $Q^u$, $R_{i}$, $R^u_{i}$.
 Then, all performance evaluation indices  can be defined
 similarly to  Section \ref{section different estimation error covariance}. Now, the nominal one-step predicted performance evaluation index $\Sigma^f_{i,k+1|k}$ and the  one-step predicted estimation  error covariance  $\Sigma^t_{i,k+1|k}$  are
defined as
\begin{equation}
\begin{aligned}
\Sigma^f_{i,k+1|k} =& \bar F_{i,k}\Sigma^f_{i,k|k-1}\bar F^T_{i,k}+ F\tilde K^f_{i,k}\tilde R^{u(L)}_{i}(F\tilde K^f_{i,k})^T+Q^u,
\end{aligned}
\end{equation}
and
\begin{equation}
\begin{aligned}
 \Sigma^t_{i,k+1|k} =&
 \bar F_{i,k}\Sigma^t_{i,k|k-1}\bar F^T_{i,k}+ F\tilde K^f_{i,k}\bar R^{(L)}_{i}(F\tilde K^f_{i,k})^T+Q,
\end{aligned}
\end{equation}
respectively, where $\bar F_{i,k} = F-F\tilde K^f_{i,k}\tilde H^{(L)}_i$ and $ \tilde K^f_{i,k}=  \Sigma^f_{i,k|k-1}(\tilde H^{(L)}_{i})^T(\tilde R^{u(L)}_{i}+\tilde H^{(L)}_{i}\Sigma^f_{i,k|k-1}(\tilde H^{(L)}_{i})^T)^{-1}$.

\vspace{6pt}
Define $\bar \Sigma^f_{i}$ and $\bar \Sigma^t_{i}$ as the solutions
to
the following discrete-time algebraic Riccati equation (DARE) and discrete-time Lyapunov equation (DLE)
\begin{equation}\label{eq riccatic t 2}
\begin{aligned}
 \Sigma^f_{i} &=
 (I-\tilde K^f_{i}\tilde H^{(L)}_i)\bar \Sigma^f_{i}(I-\tilde K^f_{i}\tilde H^{(L)}_i)^T + \tilde K^f_{i}\tilde R^{u(L)}_{i}(\tilde K^f_{i})^T,\\
  \bar \Sigma^f_{i} & =F\Sigma^f_{i}F^T+Q^u,
\end{aligned}
\end{equation}
and
\begin{equation}\label{eq riccatic t 1}
\begin{aligned}
 \Sigma^t_{i} &=
 (I-\tilde K^f_{i}\tilde H^{(L)}_i)\bar \Sigma^t_{i}(I-\tilde K^f_{i}\tilde H^{(L)}_i)^T + \tilde K^f_{i}\bar R^{(L)}_{i}(\tilde K^f_{i})^T,\\
  \bar \Sigma^t_{i} & =F\Sigma^t_{i}F^T+Q,
\end{aligned}
\end{equation}
respectively.

\vspace{6pt}
\begin{assumption}\label{ass observable}
For $i \in \mathcal{V}$,  $(F, \tilde H^{(L)}_i)$ is detectable, $(F, D)$ has no uncontrollable modes on the unit circle with $Q^u = DD^T$, and $\Sigma^f_{i,0|0}>0$.
\end{assumption}

\vspace{6pt}
\begin{theorem}\label{eq therotem convergence}
 If Assumptions  \ref{ass communication topology} and \ref{ass observable} hold,  the iterations of   $\Sigma^f_{i,k|k-1}$ and $\Sigma^t_{i,k|k-1}$ will
converge to  the solution of the DARE in (\ref{eq riccatic t 2}) and DLE in  (\ref{eq riccatic t 1}), i.e., $\bar \Sigma^f_{i}$ and $\bar \Sigma^t_{i}$, respectively.
\end{theorem}

The proof of Theorem \ref{eq therotem convergence} is given in  Appendix \ref{appdex eq therotem convergence}.

\vspace{6pt}
\begin{remark}
The collective observability condition (Assumption \ref{ass observable}) plays a crucial role in ensuring
the stability of the distributed filter. Generally, this condition is formulated  based on the actual noise covariances, however, the actual noise covariances are challenging to be obtained in practice.
Therefore,  it is essential to explore the
 convergence of the distributed filter   when
  using the collective observability condition  constructed with the nominal noise covariances.
\end{remark}

\vspace{6pt}
\begin{remark}
Theorem \ref{eq therotem convergence} shows that the convergence of the nominal performance evaluation indices can guarantee that of the estimation error covariances. Next,
since the convergence is achieved, two solutions  will be used to
evaluate the performances of the asymptotic distributed filter, and  the bounds of the asymptotic distributed   filter's performances   are  obtained by using the two equations in (\ref{eq riccatic t 2}) and (\ref{eq riccatic t 1}).
\end{remark}
%
%
%
%
%
%

\vspace{6pt}
\begin{theorem}\label{theorem sigmaf sigmat bound}
Under Assumptions \ref{ass communication topology} and  \ref{ass observable},    it follows
\begin{equation}
\begin{aligned}
 \text{max}\{0,\text{Tr}(\bar \Sigma^f_{i}) - \rho^{(L)}_i \} \leq  \text{Tr}(\bar \Sigma^t_{i})\leq \text{Tr}(\bar \Sigma^f_{i}) + \rho^{(L)}_i,
\end{aligned}
\end{equation}
where $\bar l^{(L)}_{ij} = Nl^{(L)}_{ij}(Nl^{(L)}_{ij}-1)$,
$S^f_i=F\Sigma^f_{i}P_i\Sigma^f_{i}F^T$, $P_i$ is the unique solution to  $P_i = \bar F^T_iP_i\bar F_i +I$,
\begin{equation}\label{eq theorem tr sigmaf sigmat bound}
\begin{aligned}
\rho^{(L)}_i &= \sum^{N}_{j=1}\bar l^{(L)}_{ij}||S^f_i||_f||H^T_{j}(R^{u}_{j})^{-1}H_{j} ||_f\\
&~~~+||(\tilde R^{u}_{i})^{-1}\tilde H_{i}S^f_i\tilde H^T_{i}(\tilde R^{u}_{i})^{-1}||_f||\Delta \bar R_{i}||_f\\
&~~~+||P_i||_f||\Delta Q||_f,
\end{aligned}
\end{equation}
and
\begin{equation}
\begin{aligned}
 ||\Delta \bar R_{i}||_F =\sqrt{\sum^N_{j=1}||\text{sign}(l^{(L)}_{ij})\Delta R_{j}||^2_F}.
\end{aligned}
\end{equation}

\end{theorem}
The proof of Theorem \ref{theorem sigmaf sigmat bound} is given in  Appendix \ref{appdex theorem sigmaf sigmat bound}.

\vspace{6pt}
\begin{remark}
The significance of  Theorem \ref{theorem sigmaf sigmat bound} is that the bounds of the performances of
the asymptotic distributed filter  can be evaluated  by the knowledge of the Frobenius norms of $\Delta Q$ and $\Delta R_i$.  By utilizing Proposition \ref{lemma barl approach 0},
as the fusion step  $L$ tends to infinity,
$\bar l^{(L)}_{ij}$  converges to $0$, hence,
the first term of (\ref{eq theorem tr sigmaf sigmat bound})  also converges to 0.
\end{remark}

%
%
%
%
%
%

\section{Simulations}\label{sec simulation}
\begin{figure}[!htb]
\centering
{\includegraphics[width=1.8in]{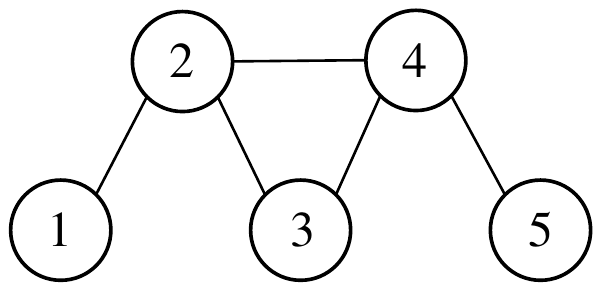}}
\caption{The diagram of the communication topology.}
\label{fig communication_topology simulation 1}
\end{figure}

In this section, the effectiveness of the theoretical results is demonstrated by utilizing some numerical simulations.
A sensor network of   five  sensors labeled from 1 to 5 is considered, and the communication topology is depicted in Fig. \ref{fig communication_topology simulation 1}.
The Metropolis weights are adopted to construct  the adjacency matrix $\mathcal{L}$ \cite{battistelli2014kullback}, described by
$l^{ij} = \frac{1}{\text{max}\{|\mathcal{N}^i|,|\mathcal{N}^j|\}}, i\in \mathcal{V}, j\in \mathcal{N}^i, i\neq j,$ and $l^{ii} = 1-\sum_{j\in\mathcal{N}^i, j\neq i}l^{ij}$.

\vspace{6pt}
To clearly  show the relations among different performance evaluation indices, the one dimensional linear system is considered. The system parameters are set as   $F = 2$, $Q=10$, $H_{i} = 1$, and $R_{i} = 10$ for all $i \in \{1,2,3,4,5\}$. Similarly to the previous definition, the standard performance evaluation index, the nominal performance evaluation index, and the estimation error covariances are denoted as $\Sigma_{i,k|k}$,
$\Sigma^f_{i,k|k}$, and $\Sigma^t_{i,k|k}$, respectively.
 Based on Assumption \ref{ass k-1 equal},  the performance evaluation indices at the previous step are chosen as  $\Sigma_{i,k-1|k-1} = \Sigma^f_{i,k-1|k-1}= \Sigma^t_{i,k-1|k-1} = 20$ for  $i \in \{1,2,3,4,5\}$.
 In addition, $\Phi_{i,k}$, $\Phi^f_{i,k}$, $\Phi^t_{i,k}$, $\Phi^{ts}_{i,k}$, and $R^{ts}_{i,k}$ are displayed
  in the simulation results to verify  the theoretical results.
  Then,  different nominal noise covariances  $R^u_{i}$ and $Q^u$ are designed for simulations.  The one-step relations  with the increasing fusion step $L$ are considered in Case 1- Case 4, and the recursive relations with the increasing time step $k$ are shown in Case 5.

\begin{figure}[!htb]
\centering
{\includegraphics[width=3.2in]{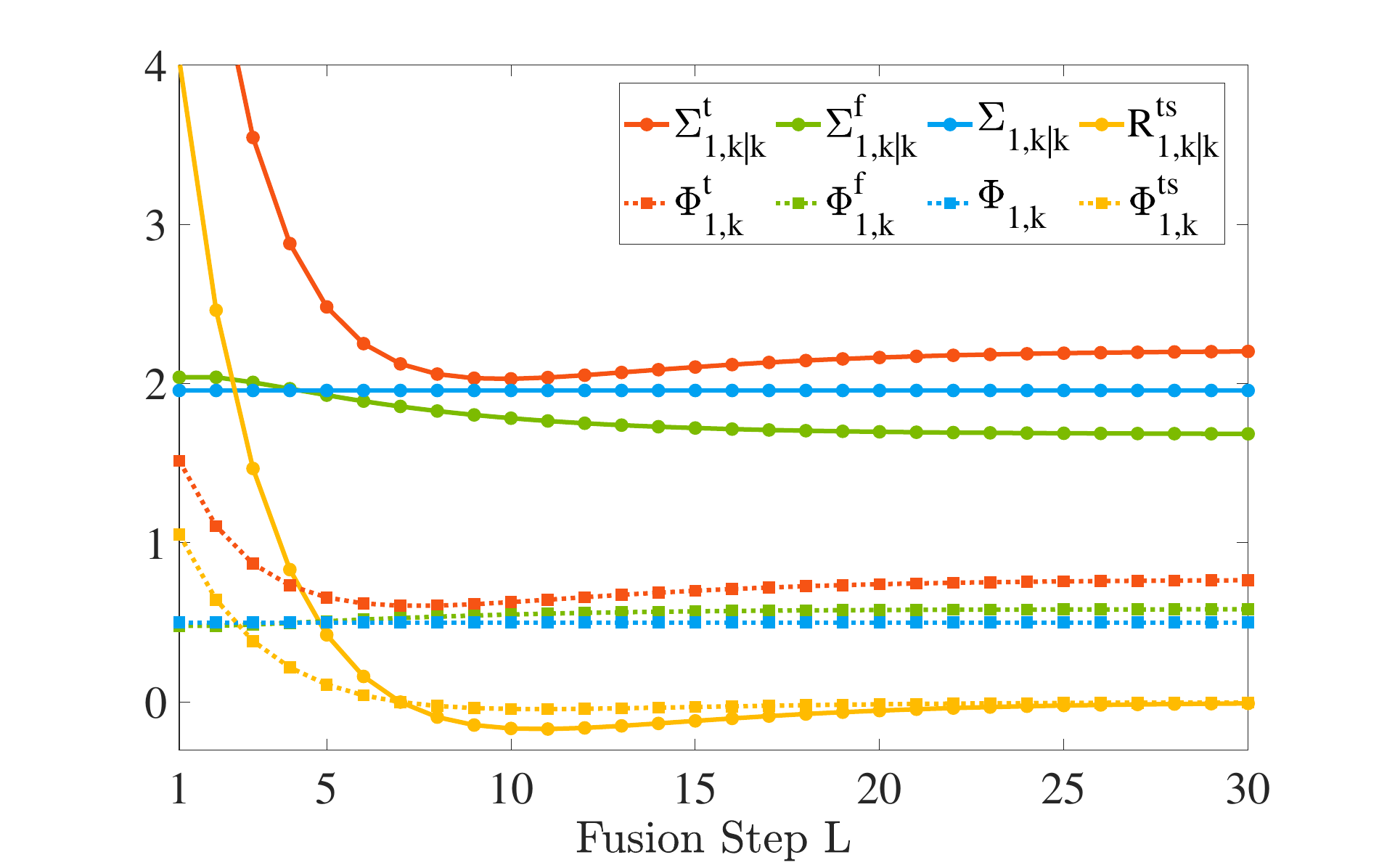}}
\caption{Illustration figure for the relations among different performance evaluation indices under $\Delta Q=0$ with the increasing fusion step $L$ in Case 1.}
\label{fig geqleq l sensor 1}
\end{figure}

\vspace{6pt}
Case 1:  Consider $\Delta Q=0$, $R^u_{i} = 10, i =1,3,4$,
$R^u_{2} = 12$, and $R^u_{5} = 5$.  Fig. \ref{fig geqleq l sensor 1} displays the relations among the different performance evaluation indices with the increasing fusion step $L$.

\begin{enumerate}
\item    When $L < 5$,  it is shown that   $\Phi^t_{i,k}>\Phi_{i,k}>\Phi^f_{i,k}$ and $\Sigma^t_{i,k|k}>\Sigma^f_{i,k|k}>\Sigma_{i,k|k}$ (See Theorem \ref{theorem phi q=0}-2).
\item When $L \geq 5$,  it can be found that $\Phi^t_{i,k}>\Phi^f_{i,k}>\Phi_{i,k}$ and $\Sigma^t_{i,k|k}>\Sigma_{i,k|k}>\Sigma^f_{i,k|k}$ (See Theorem \ref{theorem phi q=0}-1).
\end{enumerate}

\begin{figure}[!htb]
\centering
{\includegraphics[width=3.2in]{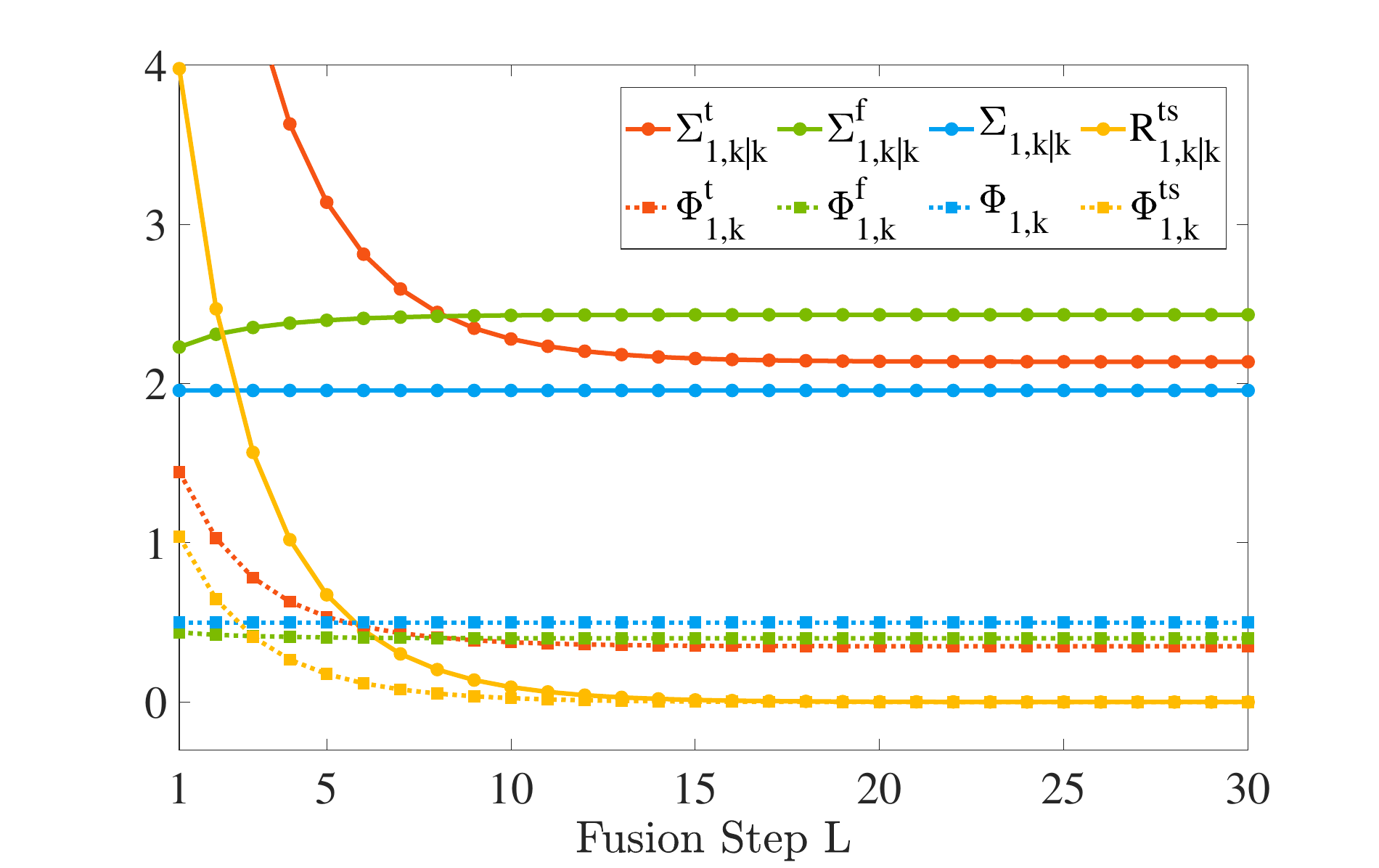}}
\caption{Illustration figure for the relations among different performance evaluation indices under $\Delta Q=0$ and $\Delta R_i\geq 0$ with the increasing fusion step $L$ in Case 2.}
\label{fig rgeq0 l sensor 1}
\end{figure}

%

\begin{figure}[!htb]
\centering
{\includegraphics[width=3.2in]{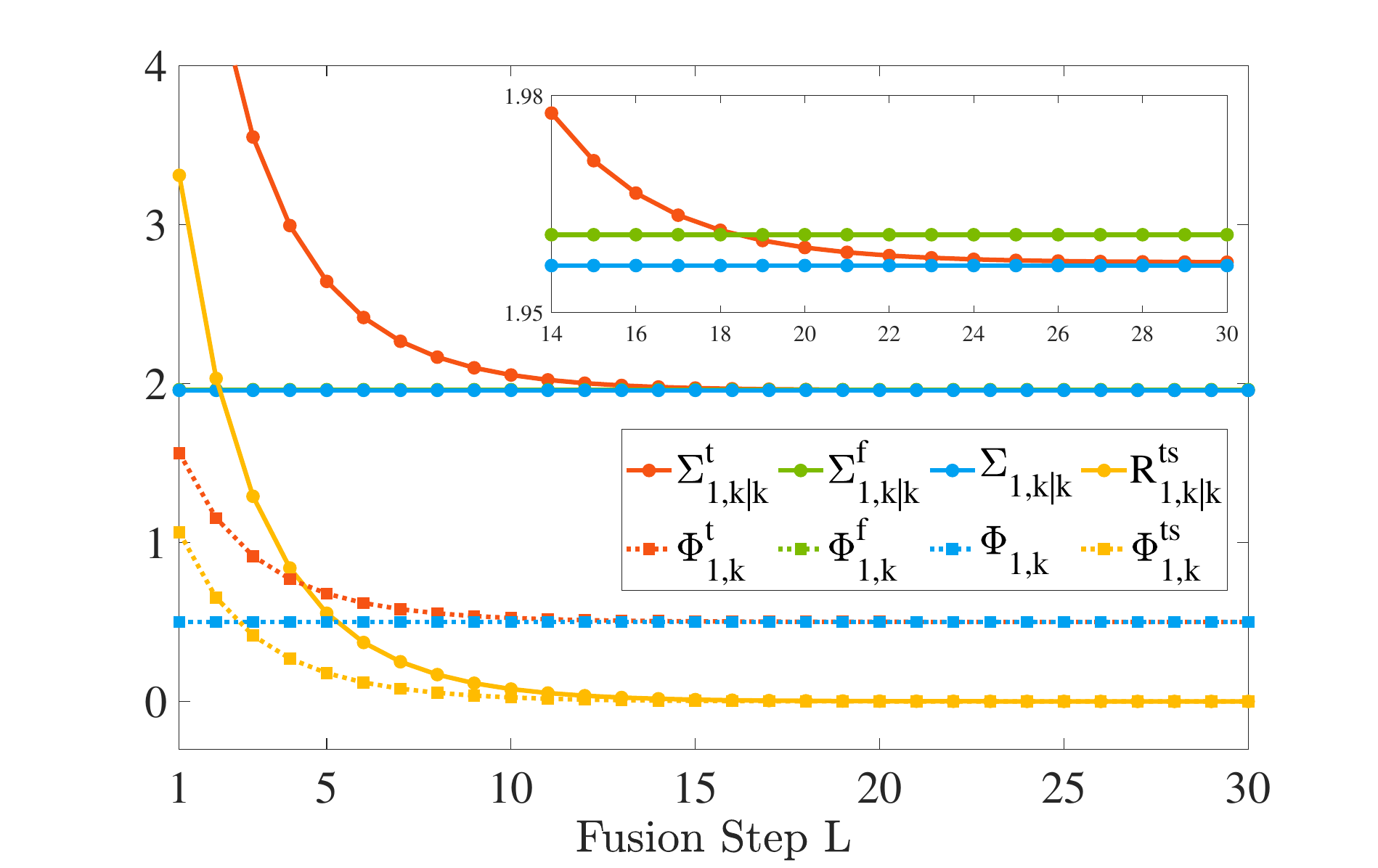}}
\caption{Illustration figure for the relations among different performance evaluation indices under $\Delta Q>0$ and $\Delta R_i= 0$ with the increasing fusion step $L$ in Case 3.}
\label{fig Qgeq0_L_sensor1}
\end{figure}

\vspace{6pt}
Case 2:  Consider $\Delta Q=0$ and $\Delta R_{i} \geq 0, i=1, 2, 3, 4, 5$. The parameters $R^u_{i}$ are selected as $R^u_{i}  = 10, i=1,3,5$  and  $R^u_{i} = 20, i=2,4$.  Fig. \ref{fig rgeq0 l sensor 1} exhibits the relations with the increasing fusion step $L$ when $\Delta Q=0$ and $\Delta R_{i} \geq 0$.

\begin{enumerate}
\item    When $L < 9$,  it is presented that   $\Phi_{i,k}>\Phi^f_{i,k}$, $\Phi^t_{i,k}>\Phi^f_{i,k}$, and $\Sigma^t_{i,k|k}>\Sigma^f_{i,k|k}>\Sigma_{i,k|k}$.
     Moreover, when $L=5$, $\Phi^t_{i,k}>\Phi_{i,k}$, and when $L=6$, $\Phi^t_{i,k}<\Phi_{i,k}$. However, it holds $\Sigma^t_{i,k|k}>\Sigma_{i,k|k}$ for all $L$, hence, it implies that the relations between $\Phi^t_{i,k}$  and $\Phi_{i,k}$  have no impact on the relations between $\Sigma^t_{i,k|k}$ and $\Sigma^f_{i,k|k}$ in this situation
    (See Theorem \ref{theorem phi q=0}-2).
\item When $L \geq 9$ and $L\to \infty$,  it holds  $\Phi_{i,k}>\Phi^f_{i,k}>\Phi^t_{i,k}$ and $\Sigma^f_{i,k|k}>\Sigma^t_{i,k|k}>\Sigma_{i,k|k}$ (See Theorem \ref{theorem deltar q=0} and Theorem \ref{theorem phi q=0}-3).
\end{enumerate}

\begin{figure}[!htb]
\centering
{\includegraphics[width=3.2in]{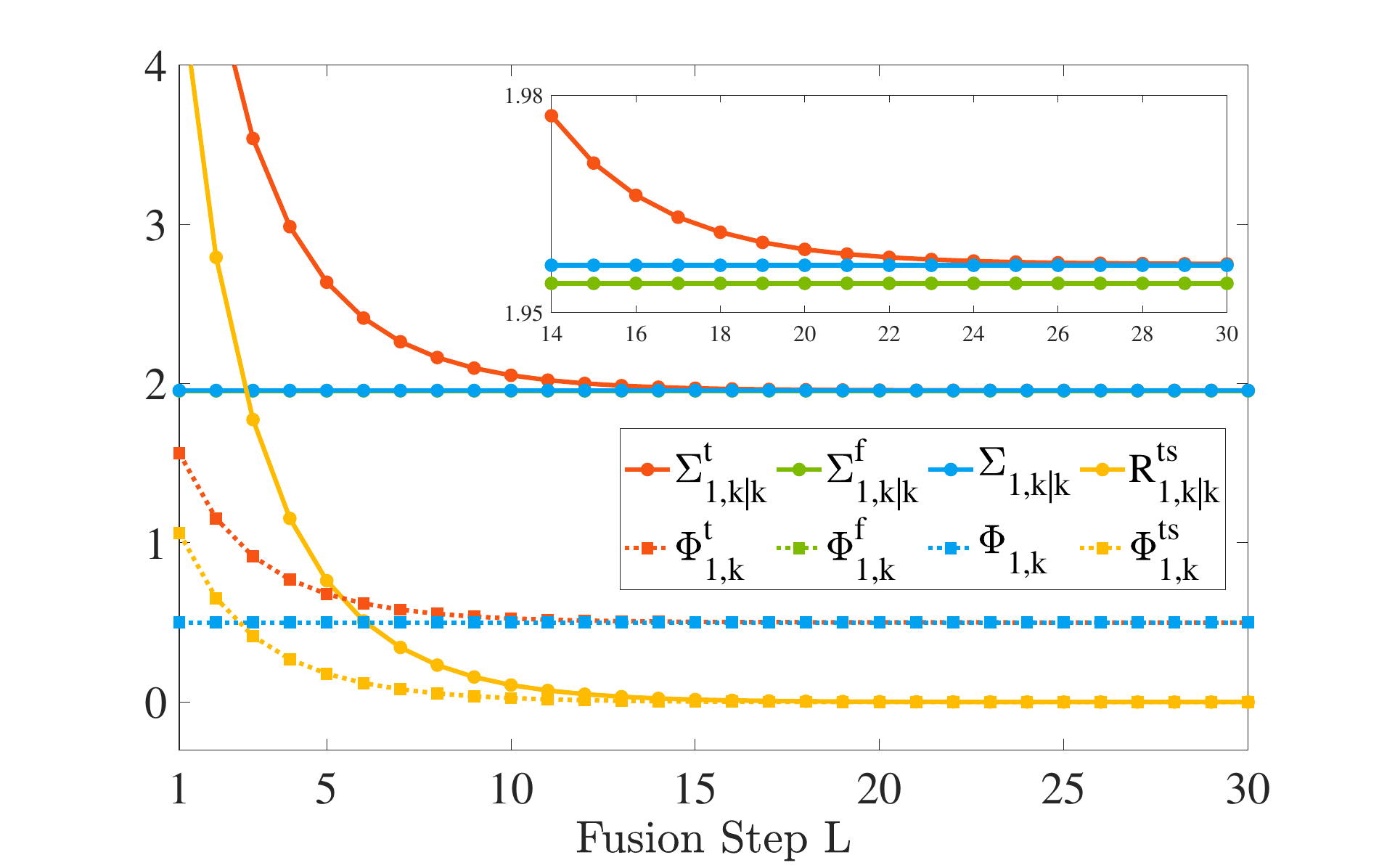}}
\caption{Illustration figure for the relations among different performance evaluation indices under $\Delta Q<0$ and $\Delta R_i= 0$ with the increasing fusion step $L$ in Case 3.}
\label{fig Qleq0_L_sensor1}
\end{figure}

\vspace{6pt}
Case 3: Consider $\Delta R_i = 0$ and $\Delta Q \neq 0$.  The nominal process covariances  are selected as $Q^u_{k} =20$ and $Q^u_{k} =5$ respectively.  Fig. \ref{fig Qgeq0_L_sensor1} and Fig. \ref{fig Qleq0_L_sensor1} illustrate the different performance evaluation indices with the increasing fusion step $L$ when $\Delta R_i = 0$, $Q^u_{k} =20$ and $\Delta R_i = 0$, $Q^u_{k} =5$,  respectively.

\begin{enumerate}
\item   From  Fig. \ref{fig Qgeq0_L_sensor1} and Fig. \ref{fig Qleq0_L_sensor1}, it can be observed that $\Phi^{t}_{i,k} = \Phi^{ts}_{i,k}-\Phi^f_{i,k}$ (See Lemma \ref{lemma phits phit-phif 1}). In addition, it is shown that $\Phi_{i,k}=\Phi^f_{i,k}$, and $\Phi^t_{i,k}$ converges to $\Phi_{i,k}$ and $\Phi^f_{i,k}$ as $L$ tends to infinity (See Lemma \ref{lemma phif=phi}).

 \item When  $\Delta Q>0$, Fig. \ref{fig Qgeq0_L_sensor1} shows that $\Sigma^f_{i,k|k}>\Sigma_{i,k|k}$. As $L$ tends to infinity,  it  holds that $\Sigma^f_{i,k|k}>\Sigma^t_{i,k|k}>\Sigma_{i,k|k}$ (See Theorem \ref{theorem r=0}-1).

\item When  $\Delta Q<0$, Fig. \ref{fig Qleq0_L_sensor1} depicts  that $\Sigma^f_{i,k|k}<\Sigma_{i,k|k}$. As $L$ tends to infinity,  it holds that $\Sigma^t_{i,k|k}>\Sigma_{i,k|k}>\Sigma^f_{i,k|k}$ (See Theorem \ref{theorem r=0}-2).
\end{enumerate}

\begin{figure}[!htb]
\centering
{\includegraphics[width=3.2in]{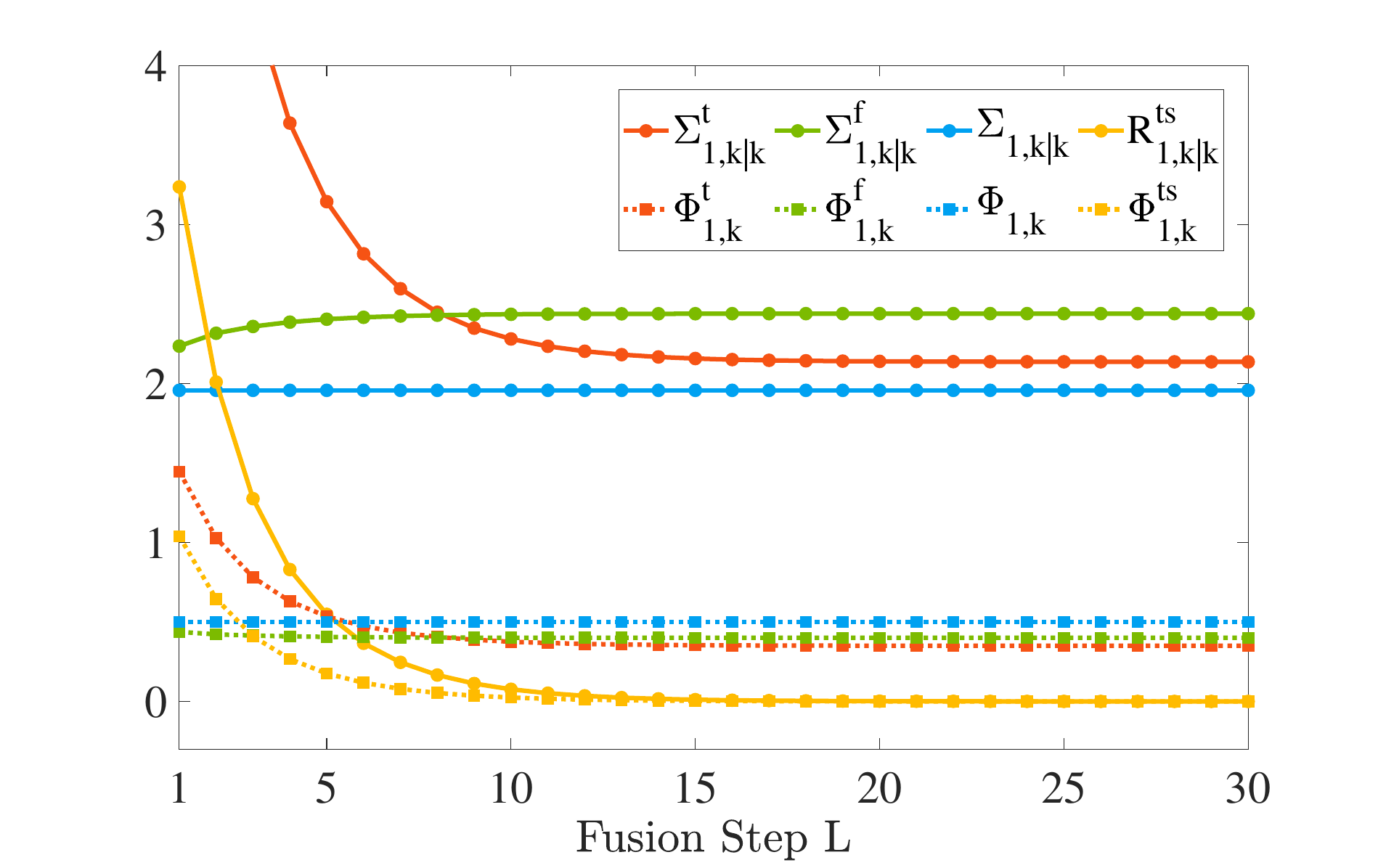}}
\caption{Illustration figure for the relations among different performance evaluation indices under $\Delta Q\neq0$ and $\Delta R_i\neq 0$ with the increasing fusion step $L$ in Case 4.}
\label{fig QR_L_sensor1}
\end{figure}

\vspace{6pt}
Case 4: Consider $\Delta R_i \geq 0$ and $\Delta Q \neq 0$.
The noise covariances are selected as $Q^u=20$, $R^u_{i}= 10, i= 1,3,5$, and $R^u_{i}=20, i = 2,4$. Then,  Fig. \ref{fig QR_L_sensor1} shows the relations among three performance evaluation indices with the increasing fusion step $L$.
\begin{enumerate}
\item When $L\geq9$, it is shown that
$\Phi^t_{i,k} \leq \Phi^f_{i,k} \leq \Phi_{i,k}$, $\Phi^{ts}_{i,k}\geq 0$, and $\Sigma_{i,k|k} \leq \Sigma^t_{i,k|k} \leq \Sigma^f_{i,k|k}$ (See Theorem \ref{theorem q r}-1).
\item  When $L<9$,  it illustrates   $\Sigma_{i,k|k} \leq \Sigma^f_{i,k|k} \leq \Sigma^t_{i,k|k}$.
\end{enumerate}

\vspace{6pt}
Overall, there are some common  characteristics from Case 1 to Case 4.
\begin{enumerate}
\item As $L$ tends to infinity,  the performance evaluation indices $\Sigma_{i,k|k}$, $\Sigma^f_{i,k|k}$, and $\Sigma^t_{i,k|k}$  converge to the stable constant values. The reason for this  phenomenon  is  that the adjacent matrix $\mathcal{L}$ converges as $L$ tends to infinity.

\item  As $L$ tends to infinity,   $\Phi^{ts}_{i,k}$  and $R^{ts}_{i,k}$ converge to zero matrix. (See Propositions  \ref{lemma sigmat-sigma} and  \ref{lemma barl approach 0}). For Case 1-4,  there exists a a tendency  $\Phi^{ts}_{i,k}\geq 0$ (See Theorem \ref{lemma katamata inequality}).
\end{enumerate}

\begin{figure}[!htb]
\centering
{\includegraphics[width=3.2in]{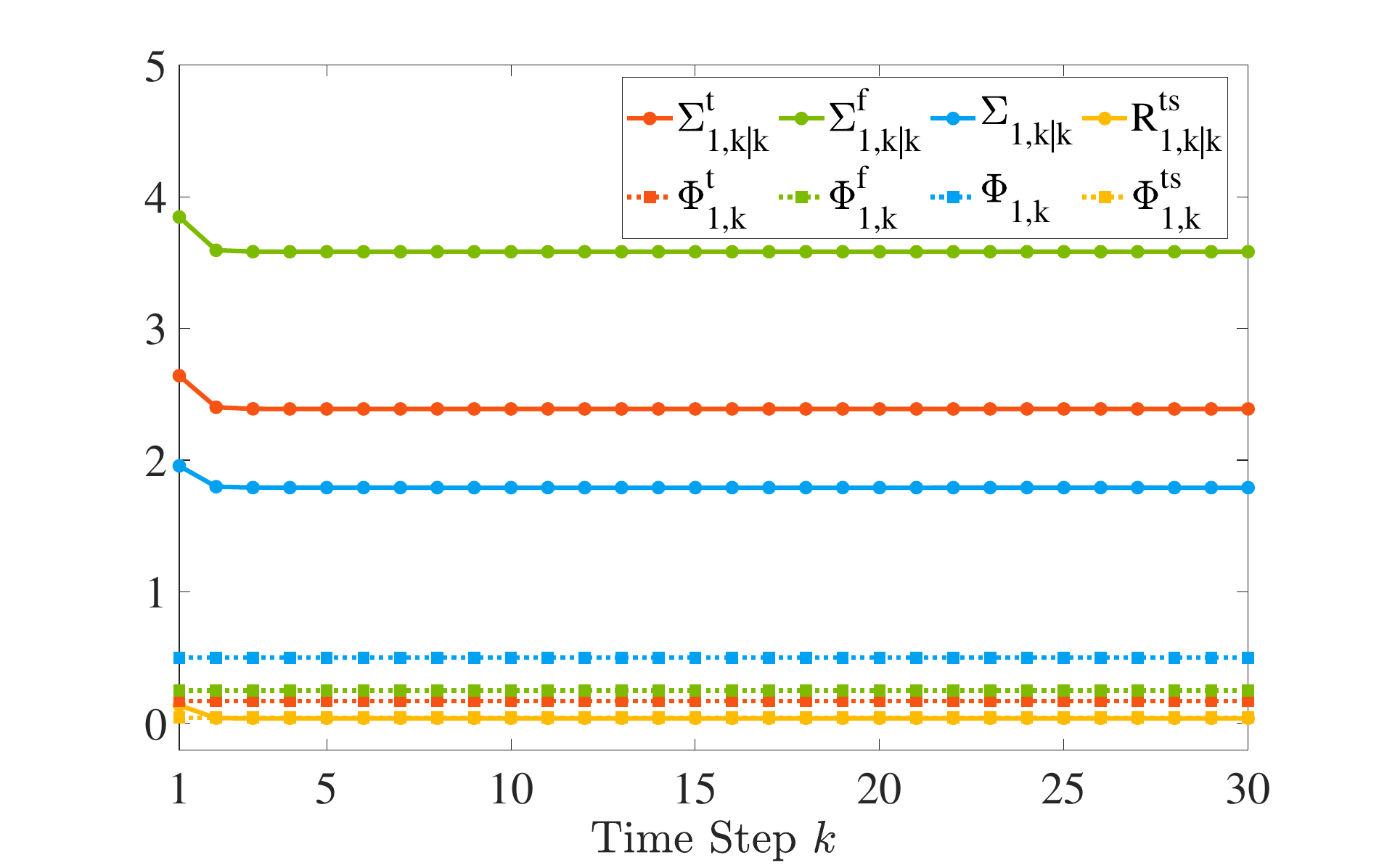}}
\caption{Illustration figure for the relations among different performance evaluation indices under $\Delta Q>0$ and $\Delta R_i> 0$ with the increasing time step $k$ in Case 5.}
\label{fig QR_k_sensor1}
\end{figure}

\vspace{6pt}
Case 5: Consider $\Delta R_i \neq 0$ and $\Delta Q \neq 0$.
The noise covariances are selected as $Q^u=20$  and $R^u_{i}=20, i = 1, 2, 3, 4, 5$. The fusion step is set as $L=5$, and the initial estimation error covariances are chosen as $\Sigma_{i,0|0} = \Sigma^f_{i,0|0}= \Sigma^t_{i,0|0} = 20$. Fig.~\ref{fig QR_k_sensor1} depicts  the relations among performance evaluation indices  with the increasing time step $k$.

\begin{enumerate}
\item For all $k>0$, it holds $\Phi_{i,k}>\Phi^f_{i,k}>\Phi^t_{i,k}$ and
 $\Sigma^f_{i,k|k} \geq \Sigma^t_{i,k|k} \geq \Sigma_{i,k|k}$ (See Theorem \ref{theorem recurse}).

\item As $k$ tends to infinity, all estimation error covariances converge to the constant values, which are the solutions of the corresponding DLEs (See Theorem \ref{eq therotem convergence}).
\end{enumerate}

\section{Conclusion}\label{sec conclusion}

In this paper, the performances of the distributed filtering under mismatched noise covariances are fully investigated from mainly two aspects.  On the one hand, the one-step and the recursive relations among three  performance evaluation indices are studied.  In addition, the effect of the consensus terms  on the one-step relations is revealed with the increasing fusion step. By extending the results of the  one-step relations,  the recursive relations are obtained with the increasing time step. On the other hand, the convergence of the nominal performance evaluation index and the estimation error covariances is derived under the convergence condition of the nominal system. In addition,  the bounds of the performances of the asymptotic distributed filter are provided by the Frobenius norm of the mismatched noise covariance deviations. These theoretical results offer  deep insights into the distributed filtering problem under mismatched noise covariances, and give powerful  instructions to the engineering applications.
In the future,  it is desired
to design the distributed filtering algorithms  under mismatched noise covariances based on these theoretical results.

%

\begin{appendices}
 \section{PROOF of Lemma \ref{lemma a+baa+b}}\label{appdex lemma 2}
 Since $aA(A+B)^{-1} = a(I+BA^{-1})^{-1}$, it can be concluded that $a(A+B)^{-1}A(A+B)^{-1}-(A+B)^{-1} =  (A+B)^{-1}(a(I+BA^{-1})^{-1}-I)$.  According to Lemma \ref{lemma matrix inverse lemma},  $a(I+BA^{-1})^{-1}-I=(a-1)I-aB(A+B)^{-1}$.

 \section{PROOF of Lemma \ref{lemma a+b-a} }\label{appdex lemma a+b-a}
Since $(A+B)^{-1}= A^{-1}-A^{-1}(B^{-1}+A^{-1})^{-1}A^{-1}$, it follows
$(A+B)^{-1}-A^{-1}= -A^{-1}(B^{-1}+A^{-1})^{-1}A^{-1}$.  In addition, note that
$-A^{-1}(AB^{-1}+I)^{-1} =  -A^{-1}B(A+B)^{-1}$, this lemma can be concluded.

\section{PROOF of Proposition \ref{lemma sigmat-sigma}}\label{appdex lemma sigmat-sigma}
By using some algebraic calculations,  (\ref{eq std sigma 22}) can be rewritten as
\begin{equation}\label{eq sigmat-sigma proof 11}
\begin{aligned}
\Sigma_{i,k|k} =& (I-\tilde K_{i,k|k}\tilde H^{(L)}_i)\Sigma_{i,k|k-1}(I-\tilde K_{i,k|k}\tilde H^{(L)}_i)^T\\
 &+ \tilde K_{i,k|k}\tilde R^{(L)}_{i,k}(\tilde K_{i,k|k})^T,
\end{aligned}
\end{equation}
where
\begin{equation*}
\begin{aligned}
\tilde K_{i,k|k}
  = \Sigma_{i,k|k-1}(\tilde H^{(L)}_{i})^T(\tilde R^{(L)}_{i,k}+\tilde H^{(L)}_{i}\Sigma_{i,k|k-1}(\tilde H^{(L)}_{i})^T)^{-1}.
\end{aligned}
\end{equation*}

Observing  (\ref{eq actual sigma2}), $\Sigma^t_{i,k|k}$ has the following equivalent form
\begin{equation}\label{eq sigmat-sigma proof 12}
\begin{aligned}
 \Sigma^t_{i,k|k} =&\tilde  \Sigma^t_{i,k|k} + R^{ts}_{i,k|k},
\end{aligned}
\end{equation}
where
\begin{equation}\label{eq sigmat-sigma proof 12-1-1-1}
\begin{aligned}
\tilde  \Sigma^t_{i,k|k} =&
 (I-\tilde K^f_{i,k}\tilde H^{(L)}_i)\Sigma^t_{i,k|k-1}(I-\tilde K^f_{i,k}\tilde H^{(L)}_i)^T~~~\\
 &+ \tilde K^f_{i,k}\tilde R^{(L)}_{i,k}(\tilde K^f_{i,k})^T,~~~~~~
\end{aligned}
\end{equation}
and
\begin{equation*}
\begin{aligned}
R^{ts}_{i,k|k} = \tilde K^f_{i,k}(\bar R^{(L)}_{i,k}-\tilde R^{(L)}_{i,k})(\tilde K^f_{i,k})^T.
\end{aligned}
\end{equation*}

 If Assumption  \ref{ass k-1 equal} satisfies, it holds  $\Sigma^t_{i,k|k-1}=\Sigma_{i,k|k-1}$ whether $\Delta Q_k\geq0$ or $\Delta Q_k\leq0$.  By utilizing   Lemma \ref{lemma k min},
it can be concluded that $\tilde  \Sigma^t_{i,k|k} - \Sigma_{i,k|k}\geq 0$, and the equality holds if and only if $\tilde R^{(L)}_{i,k}=\tilde R^{u(L)}_{i,k}$.

\vspace{6pt}
By combining (\ref{eq sigmat-sigma proof 11}) with (\ref{eq sigmat-sigma proof 12}),  $\Sigma^t_{i,k|k} - \Sigma_{i,k|k}$ can be  computed as
\begin{equation}\label{eq sigmat-sigma proof 13}
\begin{aligned}
 \Sigma^t_{i,k|k} - \Sigma_{i,k|k} &= \tilde  \Sigma^t_{i,k|k} - \Sigma_{i,k|k}+ R^{ts}_{i,k|k}.
\end{aligned}
\end{equation}

\vspace{6pt}
Next, it is desired to obtain the analysable form of   $R^{ts}_{i,k|k}$. Define the following notation
\begin{equation}\label{eq sigmat-sigma proof 14}
\begin{aligned}
M^u_{i,k}= \tilde R^{u(L)}_{i,k} +\tilde H^{(L)}_i\Sigma^f_{i,k|k-1}(\tilde H^{(L)}_i)^T.~~~~~~
\end{aligned}
\end{equation}
Hence, $R^{ts}_{i,k|k}$ can be reformulated as
\begin{equation}\label{eq sigmat-sigma proof 15}
\begin{aligned}
R^{ts}_{i,k|k}
&= \Sigma^f_{i,k|k-1}(\tilde H^{(L)}_i)^T(M^u_{i,k})^{-1}(\bar R^{(L)}_{i,k}~~~~\\
&~~~~-\tilde R^{(L)}_{i,k})(M^u_{i,k})^{-1}\tilde H^{(L)}_i\Sigma^f_{i,k|k-1}.
\end{aligned}
\end{equation}
Based on Lemma \ref{lemma matrix inverse lemma},  $(M^u_{i,k})^{-1}$ has the following form
\begin{equation}\label{eq sigmat-sigma proof 16}
\begin{aligned}
(M^u_{i,k})^{-1}&= (\tilde R^{u(L)}_{i,k})^{-1}-(\tilde R^{u(L)}_{i,k})^{-1}\tilde H^{(L)}_i((\Sigma^f_{i,k|k-1})^{-1}\\
&~~~+(\tilde H^{(L)}_i)^T(\tilde R^{u(L)}_{i,k})^{-1}\tilde H^{(L)}_i
)^{-1}(\tilde H^{(L)}_i)^T(\tilde R^{u(L)}_{i,k})^{-T}.
\end{aligned}
\end{equation}

By substituting (\ref{eq sigmat-sigma proof 16}) into (\ref{eq sigmat-sigma proof 15}) and performing some calculations, (\ref{eq sigmat-sigma proof 15}) can be rewritten as
\begin{equation*}
\begin{aligned}
 R^{ts}_{i,k|k}&= \Sigma^f_{i,k|k-1}
 \bar C_{i,k}\Phi^{ts}_{i,k}\bar C^T_{i,k}(\Sigma^f_{i,k|k-1})^T,
\end{aligned}
\end{equation*}
where
\begin{equation*}
\begin{aligned}
~~~~~\bar C_{i,k}=& I-(\tilde H^{(L)}_i)^T(\tilde R^{u(L)}_{i,k})^{-1}\tilde H^{(L)}_i((\Sigma^f_{i,k|k-1})^{-1}\\
&+(\tilde H^{(L)}_i)^T(\tilde R^{u(L)}_{i,k})^{-1}\tilde H^{(L)}_i
)^{-1},
\end{aligned}
\end{equation*}
and
\begin{equation*}
\begin{aligned}
\Phi^{ts}_{i,k}&= (\tilde H^{(L)}_i)^T(\tilde R^{u(L)}_{i,k})^{-T}
(\bar R^{(L)}_{i,k}-\tilde R^{(L)}_{i,k})
(\tilde R^{u(L)}_{i,k})^{-1}(\tilde H^{(L)}_i)\\
&= \sum^{N}_{j=1} \big((Nl^{(L)}_{ij})^2-Nl^{(L)}_{ij}\big)H^T_{j}(R^u_{j,k})^{-1}R_{j,k}(R^u_{j,k})^{-1}H_{j}.
\end{aligned}
\end{equation*}

\section{PROOF of Proposition \ref{lemma sigmat-sigmaf}}\label{appdex lemma sigmat-sigmaf}
By using the identity (\ref{eq cal identi k}), consider (\ref{eq cal sigmaf2}) and (\ref{eq actual sigma1}), and it follows
\begin{equation}
\begin{aligned}
 \Sigma^t_{i,k|k} - \Sigma^f_{i,k|k}=&\Sigma^f_{i,k|k}(\Phi^t_{i,k} - \Phi^f_{i,k})\Sigma^f_{i,k|k}+\Psi^{tf}_{i,k},
\end{aligned}
\end{equation}
where $\Psi^{tf}_{i,k}$ is defined in (\ref{eq lemma sigmat -sigmaf 1-1-1}) and
\begin{equation}\label{eq cal phit-phif}
\begin{aligned}
\Phi^t_{i,k}-\Phi^f_{i,k}
&=\sum^{N}_{j=1} Nl^{(L)}_{ij}H^T_{j}\big(Nl^{(L)}_{ij}(R^{u}_{j,k})^{-1}R_{j,k}\\~~~~~
&~~~\times(R^{u}_{j,k})^{-1}-(R^{u}_{j,k})^{-1}\big)H_{j}.~~~~~~~~~~~
\end{aligned}
\end{equation}

 By utilizing Lemma \ref{lemma a+baa+b}, (\ref{eq cal phit-phif}) can be rewritten as
\begin{equation*}
\begin{aligned}
\Phi^t_{i,k}-\Phi^f_{i,k}
&= \sum^{N}_{j=1} Nl^{(L)}_{ij}H^T_{j}(R^{u}_{j,k})^{-1}
((Nl^{(L)}_{ij}-1)I\\
&~~~~-Nl^{(L)}_{ij}\Delta R_{j,k}(R^{u}_{j,k})^{-1})H_{j}.
\end{aligned}
\end{equation*}

\section{PROOF of Proposition \ref{lemma sigmaf-sigma}}\label{appdex lemma sigmaf-sigma}
By combining  (\ref{eq std sigma post1}) and (\ref{eq cal sigmaf post1}), one has
\begin{equation}\label{eq lamma sigmaf-sigmas 1--difference}
\begin{aligned}
(\Sigma^f_{i,k|k})^{-1}- (\Sigma_{i,k|k})^{-1} = \Phi^f_{i,k}-\Phi_{i,k}+\Psi^{fs}_{i,k},
\end{aligned}
\end{equation}
where  $\Psi^{fs}_{i,k}$ is defined in (\ref{eq lemma simgf-sigmas psi}) and
\begin{equation}
\begin{aligned}
\Phi^f_{i,k}-\Phi_{i,k} =\sum^{N}_{j=1} Nl^{(L)}_{ij}H^T_{j}\big(
(R^{u}_{j,k})^{-1}-R^{-1}_{j,k}\big)H_{j}.\\
\end{aligned}
\end{equation}
Based on Lemma \ref{lemma a+b-a}, $\Phi^f_{i,k}-\Phi_{i,k}$ can be computed as
\begin{equation}
\begin{aligned}
\Phi^f_{i,k}-\Phi_{i,k}
& = \sum^{N}_{j=1} Nl^{(L)}_{ij}H^T_{j}\big(
-R^{-1}_{j,k}\Delta R_{j,k}(R^{u}_{j,k})^{-1}
\big)H_{j}.
\end{aligned}
\end{equation}

\section{PROOF of Proposition \ref{lemma single sigmat-sigma 0}}\label{appdex lemma single sigmat-sigma 0}
By making some algebraic  calculations, the proof of Proposition \ref{lemma sigmat-sigma} can be applied  to the analysis  of the filter for the single sensor.
When $N=1$, it holds  $\Phi^{ts}_{1,k}=0$ in (\ref{eq lemma sigmat-sigma 4}),  $R^{ts}_{1,k|k}=0$ and $\tilde  \Sigma^t_{1,k|k} - \Sigma_{1,k|k}\geq 0$.
 It  can be concluded that $\Sigma^t_{1,k|k}\geq \Sigma_{1,k|k}$ under any  mismatched noise covariances.

\section{PROOF of Theorem \ref{lemma barl approach 0}}\label{appdex lemma barl approach 0}
The matrix $\mathcal{\bar L}^m$  can be expressed  by using the Hadamard product ($\circ$) as follows:
\begin{equation}\label{eq lemma proof hadamard }
\begin{aligned}
\mathcal{\bar L}^m & =   N^2 \mathcal{L}^m \circ \mathcal{L}^m- N\mathcal{L}^m\circ (11^T)\\
& = N^2\mathcal{L}^m \circ (\mathcal{L}^m -\frac{1}{N} 11^T ),
\end{aligned}
\end{equation}
where the second equality uses the property $A\circ(B+C)=A\circ B+A\circ C$.
Based on Lemma \ref{lemma lapalacian metrix}\cite{qian2022consensus},  $\mathcal{L}^m$ has the one eigenvalue equal to the spectral radius 1, and the norms of all other eigenvalues are strictly less than one. Moreover,  $\mathcal{L}^m -\frac{1}{N} 11^T$  has all the eigenvalues of $\mathcal{L}^m$  except 1.  By using Lemma \ref{lemma hadamard probuct},  the spectral radius of $\mathcal{L}^m \circ (\mathcal{L}^m -\frac{1}{N} 11^T )$ in  (\ref{eq lemma proof hadamard }) can be given as
\begin{equation*}
\begin{aligned}
 \rho\big(\mathcal{L}^m \circ (\mathcal{L}^m -\frac{1}{N} 11^T )\big)&\leq \sigma\big(\mathcal{L}^m \circ (\mathcal{L}^m -\frac{1}{N} 11^T )\big)\\
 &\leq \sigma(\mathcal{L}^m) \sigma (\mathcal{L}^m -\frac{1}{N} 11^T ).
\end{aligned}
\end{equation*}

It is known that the singular values of  a Hermitian  matrix are the absolute values of the eigenvalues of this matrix. Since $\mathcal{L}$ is symmetric,  both  $\mathcal{L}^m$ and $ \mathcal{L}^m -\frac{1}{N} 11^T$ are Hermitian. Hence, one has
$\sigma(\mathcal{L}^m)\leq 1$ and  $\sigma (\mathcal{L}^m -\frac{1}{N} 11^T )< 1$. Then, it holds
\begin{equation*}
\begin{aligned}
 \rho\big(\mathcal{L}^m \circ (\mathcal{L}^m -\frac{1}{N} 11^T )\big)<1.~~~
\end{aligned}
\end{equation*}
Finally, it can be concluded that $\mathcal{\bar L}^m$ converges exponentially to $0$ as $m$ tends to infinity.

\section{PROOF of Theorem \ref{lemma katamata inequality}}\label{appdex lemma katamata inequality}
Item 1): Since  the adjacent matrix $\mathcal{L}$ is doubly stochastic,   according to  the theory of majorization \cite{marshall1979inequalities}, each column  of $\mathcal{L}^{k+1}$ is majorized by that of $\mathcal{L}^{k}$.

\vspace{6pt}
For a positive number $N$ and a  number $a$, define  the  function  $$f(x) = N^2x^2-aNx,$$
 and  it can be found that
$f(x)$ is a convex function based on the theory of  the convex function.
By using Lemma \ref{lemma majorizaiton ineuqality}, one has
$$\sum^N_{j=1}((Nl^{(m)}_{ij})^2-aNl^{(m)}_{ij})\leq \sum^N_{j=1}((Nl^{(d)}_{ij})^2-aNl^{(d)}_{ij}).$$
 Note that $l^{(m)}_{ij}$ converges to $\frac{1}{N}$ as $m\to \infty$ according to Lemma \ref{lemma lapalacian metrix}. When $a=\gamma=1$,
 it follows $\sum^N_{j=1}((Nl^{(m)}_{ij})^2-aNl^{(m)}_{ij})\geq 0$. When $a=\gamma=0$, it holds  $\sum^N_{j=1}((Nl^{(m)}_{ij})^2-aNl^{(m)}_{ij})\geq N$.
 Hence, the first part is proven.

\vspace{6pt}
Item 2): The lower bound $\underline l$ has been given in \cite{qian2021fully}. Similarly, according to the definition of majorization, the upper bound can be  also given. Since  the matrix $\mathcal{L}$ is doubly stochastic and $\sum^N_{j=1}l_{ij}=1$, the results $\bar l \geq \frac{1}{N}$ and $\underline l\leq \frac{1}{N}$ can be proven by using reduction to absurdity. Moreover, $\bar l \leq 1$ and $0\leq \underline l$ can be obtained by the properties of the graph with
 the doubly stochastic matrix $\mathcal{L}$.
 The convergence of  $\bar l$ and $\underline l$ is  derived  by using  Lemma \ref{lemma lapalacian metrix}.

 \vspace{6pt}
 Item 3): It is obvious that function $f(x) = N^2x^2-aNx$ is
   quadratic convex. If $a=1$,   the minimal value of $f(x)$ is $-\frac{1}{4}$  at the point $x=\frac{1}{2N}$.  When $x\geq \frac{1}{2N}$, the function value is monotonically increasing.
   If $a=0$,   the minimal value of $f(x)$ is $0$  at the point $x=0$. When $x\geq 0$, the function value is monotonically increasing.
   Then, the bounds of $\bar l^{(m)}_{ij}$ can be obtained  by using the properties of the quadratic function $f(x)$. Moreover, the convergence value  of $\bar l^{(m)}_{ij}$  can be given based on  Item 2) and its bounds.

\section{PROOF of Lemma \ref{lemma phito0}}\label{appdex lemma phito0}
Based on the Propositions \ref{lemma sigmat-sigma},  \ref{lemma barl approach 0},  and  Theorem \ref{lemma katamata inequality},
it can be concluded that $\Phi^{ts}_{i,k}\to 0$ as $L \to \infty$. In addition, $\bar C_{i,k}$ also converges to a fixed value, which is bounded. Hence, one has   $R^{ts}_{i,k|k}\to 0$ as $L \to \infty$.

\section{PROOF of Theorem \ref{theorem Sigmat-sigma}}\label{appdex theorem Sigmat-sigma}
Based  on  Proposition \ref{lemma sigmat-sigma}, it can be concluded that if $\Phi^{ts}_{i,k}\geq 0$, it always  holds $\Sigma^t_{i,k|k}\geq \Sigma_{i,k|k}$. The first part is proven. Then, according to the bounds of $(Nl^{(L)}_{ij})^2-Nl^{(L)}_{ij}$ (Theorem \ref{lemma katamata inequality}-3) and the convergence of $\Phi^{ts}_{i,k}$ and $R^{ts}_{i,k|k}$ (Lemma \ref{lemma phito0}), the second part can be proven.

\section{PROOF of COROLLARY \ref{corollary sigmt sigma}}\label{appdex corollary sigmt sigma}
According to  Theorem \ref{lemma katamata inequality}-1, let the number $\gamma=1$, and
multiply both sides of the inequality (\ref{eq lemma katamata ine 1}) by $H^T_{j}(R^u_{j,k})^{-1}R_{j,k}(R^u_{j,k})^{-1}H_{j}$. Then, it follows $0\leq \Phi^{ts(m)}_{i,k} \leq \Phi^{ts(d)}_{i,k}$ and  $R^{ts}_{i,k|k}\geq 0$. Then, based on  Proposition \ref{lemma sigmat-sigma}, it can be proven that $\Sigma^t_{i,k|k}\geq \Sigma_{i,k|k}$.

\section{PROOF of Lemma \ref{lemma phits phit-phif 1}}\label{appdex lemma phits phit-phif 1}
By combining (\ref{eq lemma sigmat-sigma 4}) and (\ref{eq lemma sigmat -sigmaf 1}),  the difference between  $\Phi^{ts}_{i,k}$ and $\Phi^t_{i,k}-\Phi^f_{i,k}$ can be obtained as
$$\Phi^{ts}_{i,k}- (\Phi^t_{i,k}-\Phi^f_{i,k}) =
\bar \Phi^{tf}_{i,k},$$
 where
 \begin{equation}\label{eq lemma proof sigmat-sigma}
\begin{aligned}
\bar \Phi^{tf}_{i,k} =\sum^{N}_{j=1} Nl^{(L)}_{ij}H^T_{j}(R^u_{j,k})^{-1} \Delta R_{j,k}(R^u_{j,k})^{-1}H_{j}.
\end{aligned}
\end{equation}
Then, based on  Proposition \ref{lemma barl approach 0} and Theorem \ref{lemma katamata inequality}, the limit values of $\Phi^{ts}_{i,k}$ and $\Phi^t_{i,k}-\Phi^f_{i,k}$ can be computed. Finally, according to
(\ref{eq lemma proof sigmat-sigma}) and the fact that
$\Delta R_{i,k}=0$ for all $i \in \mathcal{V}$, the last part is proven.

\section{PROOF of Theorem \ref{theorem phi q=0}}\label{appdex theorem phi q=0}
Item 1): If $\Phi^{ts}_{i,k}\geq 0$,  one has $\Sigma_{i,k|k} \leq \Sigma^t_{i,k|k}$ based on  Proposition \ref{lemma sigmat-sigma}. If $\Phi_{i,k} \leq \Phi^f_{i,k}\leq \Phi^t_{i,k}$,  it follows $ \Phi^t_{i,k}-\Phi^f_{i,k}\leq 0$ and $\Phi^f_{i,k}-\Phi_{i,k} \leq 0$.
By utilizing Propositions \ref{lemma sigmat-sigmaf} and \ref{lemma sigmaf-sigma}, it yields that $\Sigma^f_{i,k|k} \leq \Sigma_{i,k|k} \leq \Sigma^t_{i,k|k}$.

Item 2) and 3) can be derived by using the similar technique.

\section{PROOF of Theorem \ref{theorem deltar q=0}}\label{appdex theorem deltar q=0}
Item 1):  According to Lemma \ref{lemma phits phit-phif 1}, as the fusion step $L$ tends to infinity,   $\lim_{L\to \infty} \Phi^{ts}_{i,k}=0$ and $\lim_{L\to \infty} (\Phi^t_{i,k}-\Phi^f_{i,k})=-\bar \Phi^{tf}_{i,k}$. If there exists at least one $\Delta R_{i,k}>0$ and $\Delta R_{i,k}\geq 0,~\forall i\in N$,  one has  $\Phi^f_{i,k}> \Phi^t_{i,k}$ according to
$\bar \Phi^{tf}_{i,k}>0$   as $L\to \infty$. Then, it follows $\Sigma^t_{i,k|k}<\Sigma^f_{i,k|k}$. In addition,  it yields
$\Sigma_{i,k|k}<\Sigma^t_{i,k|k}$ as $L\to \infty$ by combining Proposition \ref{lemma sigmat-sigma} and Lemma \ref{lemma phits phit-phif 1}.
Moreover, by using Proposition \ref{lemma sigmaf-sigma}, one has $\Phi^f_{i,k}< \Phi_{i,k}$ and $\Sigma_{i,k|k}<\Sigma^f_{i,k|k}$.

Item 2) can be obtained by using the similar technique.

Item 3): Based on  Lemma \ref{lemma phif=phi}, it yields $\Sigma^f_{i,k|k}=\Sigma_{i,k|k}$. Then, by combining Propositions
\ref{lemma sigmat-sigma} and \ref{lemma barl approach 0}, this item can be proven.

\section{PROOF of Theorem \ref{theorem r=0}}\label{appdex theorem r=0}
Item 1): Based on  Lemma \ref{lemma phif=phi} and Proposition \ref{lemma sigmaf-sigma}, it yields $\Sigma_{i,k|k} < \Sigma^f_{i,k|k}$.  By utilizing Propositions \ref{lemma sigmat-sigma},  \ref{lemma sigmat-sigmaf} and Lemma \ref{lemma phits phit-phif 1}, this item can be proven.

Item 2) can be proven by the similar technique.

\section{PROOF of Theorem \ref{theorem recurse}}\label{appdex theorem recurse}
Item 1): The inductive method is used to prove this item. Note that
the time step begins from $k-1$ in Assumption \ref{ass k-1 equal}.
First,  when $j=k-1$, it can be concluded that  $\Sigma^f_{i,k|k} \leq \Sigma_{i,k|k} \leq \Sigma^t_{i,k|k}$ from  Theorem  \ref{theorem q r}. Then,
suppose that  it holds $\Sigma^f_{i,m-1|m-1} \leq \Sigma_{i,m-1|m-1} \leq \Sigma^t_{i,m-1|m-1}$ at time  step $j=m-1$, and we will prove $\Sigma^f_{i,m|m} \leq \Sigma_{i,m|m} \leq \Sigma^t_{i,m|m}$
at time step $j=m$. By  combining  (\ref{eq std sigma prior}),  (\ref{eq cal sigma prior}),
 (\ref{eq actual sigma prior}) with  $\Delta Q_{m-1}< 0$, it follows
 \begin{equation}\label{eq the recursive n-1 n}
\begin{aligned}
\Sigma^f_{i,m|m-1} \leq \Sigma_{i,m|m-1} \leq \Sigma^t_{i,m|m-1}.
\end{aligned}
\end{equation}

\vspace{6pt}
 For $\Sigma^f_{i,m|m}$ and $ \Sigma_{i,m|m}$,  based on (\ref{eq lamma sigmaf-sigmas 1--difference}),   one has $(\Sigma^f_{i,m|m})^{-1}- (\Sigma_{i,m|m})^{-1} = \Phi^f_{i,m}-\Phi_{i,m}+(\Sigma^f_{i,m|m-1})^{-1}- (\Sigma_{i,m|m-1})^{-1}$.  Based on   (\ref{eq the recursive n-1 n}) and $\Phi^f_{i,m}\geq \Phi_{i,m}$,  it follows  $(\Sigma^f_{i,m|m})^{-1}- (\Sigma_{i,m|m})^{-1}\geq 0$ and  $\Sigma^f_{i,m|m} \leq \Sigma_{i,m|m}$.

\vspace{6pt}
For $\Sigma^t_{i,m|m-1}$ and $ \Sigma_{i,m|m-1}$, define
\begin{equation}
\begin{aligned}
\tilde  \Sigma_{i,m|m} =&
 (I-\tilde K^f_{i,m|m}\tilde H^{(L)}_i)\Sigma_{i,m|m-1}(I-\tilde K^f_{i,m|m}\tilde H^{(L)}_i)^T\\
 &+ \tilde K^f_{i,m|m}\tilde R^{(L)}_{i,m}(\tilde K^f_{i,m|m})^T.
\end{aligned}
\end{equation}
Recalling that (\ref{eq sigmat-sigma proof 12-1-1-1}) and (\ref{eq the recursive n-1 n}), since $\Sigma_{i,m|m-1} \leq \Sigma^t_{i,m|m-1}$,
one has
$\tilde  \Sigma_{i,m|m} \leq  \tilde  \Sigma^t_{i,m|m}$.
Similarly to Proposition \ref{lemma sigmat-sigma}, it holds
$\tilde  \Sigma_{i,m|m} - \Sigma_{i,m|m}\geq 0$.
By combining  $\Phi^{ts}_{i,m}\geq 0$, it follows
$\Sigma^t_{i,m|m} - \Sigma_{i,m|m} = \tilde  \Sigma^t_{i,m|m} - \Sigma_{i,m|m}+ R^{ts}_{i,m|m}\geq \tilde  \Sigma_{i,m|m} - \Sigma_{i,m|m}+ R^{ts}_{i,m|m}\geq 0$ and $\Sigma^t_{i,m|m} \geq \Sigma_{i,m|m}$.

\vspace{6pt}
For $\Sigma^t_{i,m|m-1}$ and $ \Sigma^f_{i,m|m-1}$, since $\Phi_{i,m}\leq \Phi^f_{i,m}\leq \Phi^t_{i,m}$, by combining
(\ref{eq lemma sigmat -sigmaf 1})  and  (\ref{eq lemma sigmat -sigmaf 1-1-1}),  it holds  $\Sigma^t_{i,m|m} \geq  \Sigma^f_{i,m|m}$.
Based on the above analysis,  it follows  $\Sigma^f_{i,m|m} \leq \Sigma_{i,m|m} \leq \Sigma^t_{i,m|m}$, and the proof can be concluded.

 Item 2) can be proven by utilizing the similar technique.

\section{PROOF of Theorem \ref{eq therotem convergence}}\label{appdex eq therotem convergence}
Based on  Assumption \ref{ass observable},  it can be concluded that  $\Sigma^f_{i,k|k-1}$ converges to $\bar \Sigma^f_{i}$, and $\bar F_i = F(I-\tilde K^f_{i}\tilde H^{(L)}_i)$ is Schur stable \cite{chan1984convergence}. As $k$ tends to infinity,  it follows
 $\lim_{k\to \infty}\tilde K^f_{i,k} = \tilde K^f_{i}$ and
  $\lim_{k\to \infty}\bar F_{i,k} = \bar F_{i}$.
By using Theorem 1 in \cite{cattivelli2010diffusion,qian2022consensus} and the fact that $\bar F_i = F(I-\tilde K^f_{i}\tilde H^{(L)}_i)$ is Schur stable, it can be proven that $\Sigma^t_{i,k|k-1}$ converges to $\bar \Sigma^t_{i}$.

\section{PROOF of Theorem \ref{theorem sigmaf sigmat bound}}\label{appdex theorem sigmaf sigmat bound}
First, by using some  algebraic operation,   $\bar \Sigma^t_i$ can be expressed as
\begin{equation}\label{eq secalg Sigmat trans}
\begin{aligned}
\bar \Sigma^t_{i} =&  F\Sigma^f_{i}(\bar\Sigma^f_{i})^{-1}\bar\Sigma^t_{i}(\bar\Sigma^f_{i})^{-1}\Sigma^f_{i}F^T+F\Sigma^f_{i} (\tilde H^{(L)}_{i})^T\\
&\times(\tilde R^{u(L)}_{i})^{-1} \bar R^{(L)}_{i}(\tilde R^{u(L)}_{i})^{-1}\tilde H^{(L)}_{i}\Sigma^f_{i}F^T +Q.
\end{aligned}
\end{equation}

Then, consider the vectorization of (\ref{eq secalg Sigmat trans}). By utilizing the identity  $\text{vec}(ABC) = (C^T\otimes A)\text{vec}(B)$, it follows
\begin{equation}\label{eq secalg vecsigmat 1}
\begin{aligned}
\text{vec}(\bar \Sigma^t_{i}) =&  (\bar F_i\otimes \bar F_i) \text{vec}(\bar\Sigma^t_{i})+((F\Sigma^f_{i})\otimes(F\Sigma^f_{i}))\text{vec}((\tilde H^{(L)}_{i})^T\\
&\times(\tilde R^{u(L)}_{i})^{-1} \bar R^{(L)}_{i}(\tilde R^{u(L)}_{i})^{-1}\tilde H^{(L)}_{i})+\text{vec}(Q),
\end{aligned}
\end{equation}
where $\bar F_i = F\Sigma^f_{i}(\bar\Sigma^f_{i})^{-1}$.  Let $T_i = I - \bar F_i\otimes \bar F_i$.  In virtue of (\ref{eq cal identi i-kh}),  it holds  $\Sigma^f_{i}(\bar\Sigma^f_{i})^{-1} = I-\tilde K^f_{i}\tilde H^{(L)}_{i}$ and   $\bar F_i = F-F\tilde K^f_{i}\tilde H^{(L)}_{i}$. In addition, $\bar F_i$ is Schur stable based on the proof  of Theorem \ref{eq therotem convergence}. Hence, it can be concluded that $T_i$  is also Schur stable according to the spectrum  property the Kronecker product. Next, (\ref{eq secalg vecsigmat 1}) can  be computed as
 \begin{equation}\label{eq secalg vecsigmat 2}
\begin{aligned}
\text{vec}(\bar \Sigma^t_{i}) &=  T^{-1}_i((F\Sigma^f_{i})\otimes(F\Sigma^f_{i}))\text{vec}((\tilde H^{(L)}_{i})^T(\tilde R^{u(L)}_{i})^{-1}\\
 &~~~\times \bar R^{(L)}_{i}(\tilde R^{u(L)}_{i})^{-1}\tilde H^{(L)}_{i})+ T^{-1}_i\text{vec}(Q).~~~~~~~~~~~
\end{aligned}
\end{equation}

By using the identity $\text{Tr}(A) = (\text{vec}(I))^T \text{vec}(A)$, multiply both sides of the equation (\ref{eq secalg vecsigmat 2}) by $(\text{vec}(I))^T$, and it follows
\begin{equation}\label{eq secalg vecsigmat 3}
\begin{aligned}
\text{Tr}(\bar \Sigma^t_{i}) =& (\text{vec}(P_i))^T((F\Sigma^f_{i})\otimes(F\Sigma^f_{i}))\text{vec}((\tilde H^{(L)}_{i})^T
(\tilde R^{u(L)}_{i})^{-1}\\
&\times \bar R^{(L)}_{i}(\tilde R^{u(L)}_{i})^{-1}\tilde H^{(L)}_{i})+ (\text{vec}(P_i))^T\text{vec}(Q),
\end{aligned}
\end{equation}
where $(\text{vec}(P_i))^T = (\text{vec}(I))^T T^{-1}_i$. By performing the similar algebraic calculations,  $P_i$ can be obtained by solving    $P_i = \bar F^T_iP_i\bar F_i +I$, and $P_i$ is the unique solution.
Then,  (\ref{eq secalg vecsigmat 3}) is rewritten as
\begin{equation}\label{eq secalg vecsigmat 4}
\begin{aligned}
\text{Tr}(\bar \Sigma^t_{i})
& = (\text{vec}(F\Sigma^f_{i}P_i\Sigma^f_{i}F^T))^T\text{vec}((\tilde H^{(L)}_{i})^T
(\tilde R^{u(L)}_{i})^{-1}\\
&~~~\times\bar R^{(L)}_{i}(\tilde R^{u(L)}_{i})^{-1}\tilde H^{(L)}_{i})+ (\text{vec}(P_i))^T\text{vec}(Q).~~~~
\end{aligned}
\end{equation}

Like (\ref{eq secalg Sigmat trans})-(\ref{eq secalg vecsigmat 4}),  $\text{Tr}(\bar \Sigma^f_{i})$  can be given as
\begin{equation}\label{eq secalg vecsigmat 5}
\begin{aligned}
\text{Tr}(\bar \Sigma^f_{i}) &= (\text{vec}(F\Sigma^f_{i}P_i\Sigma^f_{i}F^T))^T\text{vec}((\tilde H^{(L)}_{i})^T\\
&~~~\times(\tilde R^{u(L)}_{i})^{-1} \tilde H^{(L)}_{i})+(\text{vec}(P_i))^T\text{vec}(Q^u).~~~~~~~
\end{aligned}
\end{equation}

%
%
%

By combining (\ref{eq secalg vecsigmat 4}) and (\ref{eq secalg vecsigmat 5}),  it holds
\begin{equation}\label{eq secalg vecsigmat 6}
\begin{aligned}
\text{Tr}(\bar \Sigma^t_{i}) -  \text{Tr}(\bar \Sigma^f_{i}) &= (\text{vec}(F\Sigma^f_{i}P_i\Sigma^f_{i}F^T))^T\Phi^{\text{vec}}_i~~~~~~~~\\
&~~~+(\text{vec}(P_i))^T\text{vec}(Q^u),
\end{aligned}
\end{equation}
where
\begin{equation}\label{eq secalg vecsigmat 7}
\begin{aligned}
 \Phi^{\text{vec}}_i & = \text{vec}((\tilde H^{(L)}_{i})^T(\tilde R^{u(L)}_{i})^{-1} \bar R^{(L)}_{i}(\tilde R^{u(L)}_{i})^{-1}\tilde H^{(L)}_{i})\\
&~~~~ -\text{vec}((\tilde H^{(L)}_{i})^T(\tilde R^{u(L)}_{i})^{-1} \tilde H^{(L)}_{i}).
\end{aligned}
\end{equation}

According to Proposition \ref{lemma sigmat-sigmaf}, (\ref{eq secalg vecsigmat 7}) can be rewritten as
\begin{equation}\label{eq secalg vecsigmat 8}
\begin{aligned}
 \Phi^{\text{vec}}_i & = \sum^N_{j=1}  \bar l^{(L)}_{ij}\text{vec}\big(H^T_{j}(R^{u}_{j})^{-1}H_{j} )\\
&~~~-\sum^{N}_{j=1} (Nl^{(L)}_{ij})^2\text{vec}(H^T_{j}(R^{u}_{j})^{-1}\Delta R_{j}(R^{u}_{j})^{-1}H_{j}\big).
\end{aligned}
\end{equation}

The second term  in (\ref{eq secalg vecsigmat 8}) has the following form by using the identity  $\text{vec}(ABC) = (C^T\otimes A)\text{vec}(B)$
\begin{equation}
\begin{aligned}
&~~~~ \sum^{N}_{j=1} (Nl^{(L)}_{ij})^2\text{vec}(H^T_{j}(R^{u}_{j})^{-1}\Delta R_{j}(R^{u}_{j})^{-1}H_{j})\\
& = \text{vec}(\tilde H^T_{i}(\tilde R^{u}_{i})^{-1} \Delta \bar R_{i}( \tilde R^{u}_{i})^{-1}\tilde H_{i})\\
& = ((\tilde H^T_{i}(\tilde R^{u}_{i})^{-1})\otimes (\tilde H^T_{i}(\tilde R^{u}_{i})^{-1}))\text{vec}( \Delta \bar R_{i}).~~
\end{aligned}
\end{equation}

Define $S^f_i=F\Sigma^f_{i}P_i\Sigma^f_{i}F^T$, and it holds
\begin{equation}
\begin{aligned}
&~~~(\text{vec}(S^f_i))^T((\tilde H^T_{i}(\tilde R^{u}_{i})^{-1})\otimes (\tilde H^T_{i}(\tilde R^{u}_{i})^{-1}))\text{vec}( \Delta \bar R_{i})\\
 &=  (\text{vec}((\tilde R^{u}_{i})^{-1}\tilde H_{i}S^f_i\tilde H^T_{i}(\tilde R^{u}_{i})^{-1}))^T\text{vec}( \Delta \bar R_{i}).
\end{aligned}
\end{equation}

Next, (\ref{eq secalg vecsigmat 6}) can be rewritten as
\begin{equation}
\begin{aligned}
\text{Tr}(\bar \Sigma^t_{i})-\text{Tr}(\bar \Sigma^f_{i}) =  \rho^b_i,
\end{aligned}
\end{equation}
where
\begin{equation}
\begin{aligned}
\rho^b_i &= \sum^N_{j=1}  \bar l^{(L)}_{ij}(\text{vec}(S^f_i))^T\text{vec}(H^T_{j}(R^{u}_{j})^{-1}H_{j} )\\
&~~~+ (\text{vec}((\tilde R^{u}_{i})^{-1}\tilde H_{i}S^f_i\tilde H^T_{i}(\tilde R^{u}_{i})^{-1}))^T\text{vec}( \Delta \bar R_{i})\\
&~~~-(\text{vec}(P_i))^T\text{vec}(\Delta Q).
\end{aligned}
\end{equation}

Since $||\text{vec}(A)||_F=||A||_F$, it follows  $||\rho^b_i ||_F \leq \rho^{(L)}_i$, where
\begin{equation}
\begin{aligned}
\rho^{(L)}_i &= \sum^{N}_{j=1}\bar l^{(L)}_{ij}||S^f_i||_F||H^T_{j}(R^{u}_{j})^{-1}H_{j} ||_F\\
&~~~+||(\tilde R^{u}_{i})^{-1}\tilde H_{i}S^f_i\tilde H^T_{i}(\tilde R^{u}_{i})^{-1}||_F||\Delta \bar R_{i}||_F ~~\\
&~~~+||P_i||_F||\Delta Q||_F,
\end{aligned}
\end{equation}
and
\begin{equation}
\begin{aligned}
 ||\Delta \bar R_{i}||_F =\sqrt{\sum^N_{j=1}||\text{sign}(l^{(L)}_{ij})\Delta R_{j}||^2_F}.
\end{aligned}
\end{equation}
Since $\bar \Sigma^t_{i}$ is non-negative definite, the lower bound of $\text{Tr}(\bar \Sigma^t_{i})$  is useful
only when $\text{Tr}(\bar \Sigma^f_{i}) \geq \rho^{(L)}$, and $\text{Tr}(\bar \Sigma^t_{i})\geq 0$.

\end{appendices}

\bibliographystyle{IEEEtran}
\bibliography{ref}

\end{document}